\RequirePackage{graphicx}

\documentclass[review, 3p]{elsarticle}
\usepackage{lineno,hyperref}
\usepackage{mathtools}
\usepackage{amssymb}
\usepackage{multirow}
\usepackage[pdftex]{color}
\usepackage{caption}
\usepackage{subcaption}
\usepackage{amsthm}

\usepackage{cleveref}

\usepackage{xpatch}
\xpatchcmd{\MaketitleBox}{\hrule}{}{}{}
\xpatchcmd{\MaketitleBox}{\hrule}{}{}{}

\makeatletter
\renewcommand{\fnum@figure}{Fig. \thefigure}
\makeatother

\newtheorem{theorem}{Theorem}
\newdefinition{remark}{Remark}
\newtheorem{conjecture}{Conjecture}
\newtheorem{corollary}{Corollary}
\newtheorem{proposition}{Proposition}
\newtheorem{lemma}{Lemma}
\newdefinition{definition}{Definition}

\usepackage{hyphenat}

\DeclarePairedDelimiter\floor{\lfloor}{\rfloor}

\newproof{proofof}{\textbf{Proof}}

\modulolinenumbers[5]

\biboptions{sort&compress}

\journal{Journal of Multivariate Analysis}

\makeatletter
\def\ps@pprintTitle{%
  \let\@oddhead\@empty
  \let\@evenhead\@empty
  \let\@oddfoot\@empty
  \let\@evenfoot\@oddfoot
}
\makeatother

\begin{document}

\begin{frontmatter}


\title{Extreme Eigenvalue Distributions of Jacobi Ensembles:\\ New Exact Representations, Asymptotics and Finite Size Corrections}







\author{Laureano Moreno-Pozas\corref{mycorrespondingauthor}}
\cortext[mycorrespondingauthor]{ECE Department, Hong Kong University of Science and Technology, Hong Kong; emails: eelaureano@ust.hk, m.mckay@ust.hk\sloppy}

\author{David Morales-Jimenez\corref{mycorrespondingauthor2}}
\cortext[mycorrespondingauthor2]{ECIT Institute, Queen’s University Belfast, UK; email: d.morales@qub.ac.uk\\ This work was supported by the HKTIIT (HKTIIT16EG01) and the General Research Fund of the Hong Kong Research Grants Council (16202918).}

\author{Matthew R. McKay\corref{mycorrespondingauthor}}





\begin{abstract}
Let $\mathbf{W}_1$ and $\mathbf{W}_2$ be independent $n\times n$ complex central Wishart matrices with $m_1$ and $m_2$ degrees of freedom respectively. This paper is concerned with the extreme eigenvalue distributions of double-Wishart matrices $(\mathbf{W}_1+\mathbf{W}_2)^{-1}\mathbf{W}_1$, which are analogous to those of F matrices ${\bf W}_1 {\bf W}_2^{-1}$ and those of the Jacobi unitary ensemble (JUE). Defining $\alpha_1=m_1-n$ and $\alpha_2=m_2-n$, we derive new exact distribution formulas in terms of $(\alpha_1+\alpha_2)$-dimensional matrix determinants, with elements involving derivatives of Legendre polynomials.  This provides a convenient exact representation, while facilitating a direct large-$n$ analysis with $\alpha_1$ and $\alpha_2$ fixed (i.e., under the so-called ``hard-edge" scaling limit); the analysis is based on new asymptotic properties of Legendre polynomials and their relation with Bessel functions that are here established. Specifically, we present limiting formulas for the smallest and largest eigenvalue distributions as $n \to \infty$ in terms of $\alpha_1$- and $\alpha_2$-dimensional determinants respectively, which agrees with expectations from known universality results involving the JUE and the Laguerre unitary ensemble (LUE). We also derive finite-$n$ corrections for the asymptotic extreme eigenvalue distributions under hard-edge scaling, giving new insights on universality by comparing with corresponding correction terms derived recently for the LUE.  Our derivations are based on elementary algebraic manipulations, differing from existing results on double-Wishart and related models which often involve Fredholm determinants, Painlev\'e differential equations, or hypergeometric functions of matrix arguments.
\end{abstract}


\end{frontmatter}

\linenumbers

\section{Introduction}





Double Wishart random matrices, defined as ${\mathbf{W}=(\mathbf{W}_1+\mathbf{W}_2)^{-1}\mathbf{W}_1}$, with $\mathbf{W}_1$ and $\mathbf{W}_2$ $n\times n$ Wishart with $m_1$ and $m_2$ degrees of freedom respectively, are an important class of random matrix models. They find application in multivariate analysis of variance (MANOVA), where corresponding test statistics involve the eigenvalues of $\mathbf{W}$, either the complete set or simply the extreme largest/smallest eigenvalues \cite{Roy1953heuristic}. For linear hypothesis testing, the natural ``null hypothesis'' considers $\mathbf{W}_1$ and $\mathbf{W}_2$ independent, central, having identical covariance matrix.  The eigenvalues of $\mathbf{W}$ in this case are intimately connected with those of classical Jacobi ensembles and those of Fisher (or F) matrices $\mathbf{F}=\mathbf{W}_1 \mathbf{W}_2^{-1}$, by appropriate variable transformations. Here, we present new results for the extreme eigenvalues of ${\bf W}$ for the case of $\mathbf{W}_1$ and $\mathbf{W}_2$ being complex Wishart, hence yielding analogous results for the classical Jacobi unitary ensemble (JUE) and complex F model.  In addition to their use in statistical testing, the extreme eigenvalues of such complex models arise in multi-antenna communication systems with co-channel interference \cite{Kang2004quadratic} and in quantum conductance in mesoscopic physics \cite{Beenakker1997, Forrester2006quantum}.

Our results further contribute to a large amount of prior work on the extreme eigenvalues of double Wishart models (equivalently, JUE/F models).  Exact expressions for the extreme eigenvalue distributions of ${\bf W}$ have been given in terms of Fredholm determinants \cite{Tracy1994fredholm, Johnstone2008}, or through equivalent representations in terms of solutions to Painlev\'e differential equations \cite{Tracy1994fredholm, Haine1999jacobi}. Other exact results have been given in terms of \mbox{$n$-dimensional determinants} \cite{Khatri1}, Gauss hypergeometric functions \cite{Borodin2003}, and polynomial expansions involving combinatorial sums \cite{Borodin2003, Dumitriu2008}. These results are summarized in \mbox{Table \ref{table_1}}. 

The asymptotic distributions of the extreme eigenvalues of $\mathbf{W}$ have also been studied as $n$, $m_1$, and $m_2$ become large.  
Particularly noteworthy are results obtained by taking asymptotics on the Fredholm determinant representation \cite{Tracy1994fredholm, Johnstone2008}, a determinant expansion involving the so-called Jacobi kernel \cite{Johnstone2008}. 
This kernel has been shown to converge to the well-known Bessel kernel \cite{Nagao1993}, when appropriately scaled under the ``hard-edge'' scaling regime, $n\rightarrow\infty$ with $\alpha_1=m_1-n$ and $\alpha_2=m_2-n$ fixed \cite{Borodin2003, Perret2016, Forrester2019finite}. 
Consequently, under hard-edge asymptotics, it follows that the extreme eigenvalue distributions of $\mathbf{W}$ can be expressed in terms of a Fredholm determinant involving the Bessel kernel, which has also been shown to admit an equivalent integral form involving the solution of a Painlev\'e III differential equation \cite{TW1994}.
Along a different line, hard-edge asymptotics were evaluated directly in \cite{Borodin2003}, based on the exact hypergeometric function representation, where the smallest eigenvalue distribution of $\mathbf{W}$ was shown to be expressible in terms of a Bessel hypergeometric function of a $\alpha_1$-dimensional matrix argument.

It is important to note that an analogous Fredholm determinant representation involving the Bessel kernel has also been established for the smallest eigenvalue distribution of the Laguerre unitary ensemble (LUE) under a similar hard-edge scaling limit \cite{Forrester1993}, suggesting a form of universality among the behavior of the smallest eigenvalue of the LUE and that of the extreme eigenvalues of ${\bf W}$ under hard-edge asymptotics. Remarkably, in the context of the LUE 
, this asymptotic distribution was shown to admit a very simple representation in terms of a finite-dimensional determinant involving Bessel functions \cite{Forrester1994alpha_det}.  Such a representation should also apply for the extreme eigenvalues of ${\bf W}$, though this has not been explicitly shown.

In turn, there exists another asymptotic regime, referred to as the ``soft-edge'' scaling regime, for which $n\rightarrow\infty$ and either $\alpha_1\rightarrow\infty$ or $\alpha_2\rightarrow\infty$. Under this regime, the Jacobi kernel was shown to converge to the Airy kernel \cite{Borodin2003}; thus, the extreme eigenvalue distributions of $\mathbf{W}$ can be expressed in terms of a Fredholm determinant involving the Airy kernel. Alternatively, an integral form has been established, involving the solution of a Painlev\'e II differential equation \cite{Tracy-Widom1994}, which is the widely recognized Tracy-Widom law. Also noteworthy is the fact that an analogous representation involving the Airy kernel has been established for the smallest eigenvalue of the LUE \cite{Tracy1994fredholm}, so that the said form of universality between the extreme eigenvalues of the JUE and the LUE holds more generally, not only under hard-edge, but also under soft-edge asymptotics. Such universality has been suggested to hold even more generally, for other statistics beyond the extreme eigenvalue distributions \cite{ForresterBookLoggases}.

\begin{table}[t!]
\renewcommand{\arraystretch}{1.7}
\caption{Previous exact results for the extreme eigenvalue distributions of complex Jacobi ensembles.}
\label{table_1}
\centering
\begin{tabular}
{c|c}
\hline
\text{1. $n$-dimensional determinant}  & \cite{Khatri1}  \\
\text{2. Fredholm determinant} &  \cite{Tracy1994fredholm, Johnstone2008}\\
\text{3.  Gauss hypergeometric function of $\alpha_1$ or $\alpha_2$-dimensional argument}  &  \cite{Borodin2003}\\
\text{4. Painlev\'e VI} &  \cite{Haine1999jacobi}\\
\text{5. Polynomial of degree $n\alpha_1$ or $n\alpha_2$ involving partitions}  & \cite{Borodin2003,Dumitriu2008}\\
\hline
\end{tabular}
\end{table}



Despite the extensive literature regarding the JUE and LUE, results are scarce when one considers departure from universality. In particular,  Edelman, Guionnet and P\'ech\'e have recently conjectured a first-order correction proportional to $n^{-1}$ for the smallest eigenvalue distribution of the LUE under the hard-edge scaling limit.
Independent proofs for this correction have been provided by Perret and Schehr \cite{Perret2016} and Bornemann \cite{Bornemann2016}. In a very recent unpublished manuscript \cite{Forrester2019finite}, Forrester and Trinh studied the optimal scaling for the smallest eigenvalue distribution of the Laguerre $\beta$-ensemble, which subsumes the real ($\beta=1$), complex ($\beta=2$) and symplectic ($\beta=4$) cases, and provide a first-order correction proportional to $n^{-2}$ for $\beta=2$ in integral form, involving a solution to a second-order differential equation. However, to the best of our knowledge, finite-$n$ corrections for the extreme eigenvalue distributions of $\mathbf{W}$ (or the JUE) are not available thus far. A question remains as to whether the universality between LUE and JUE persists when considering finite-$n$ corrections?

A goal of this paper is to provide finite-$n$ corrections for the extreme eigenvalue distributions of $\mathbf{W}$, and therefore, for the F model and the classical JUE, under hard-edge asymptotics. By exploiting new exact representations of the extreme eigenvalue distributions of $\mathbf{W}$, we perform an asymptotic analysis under the hard-edge scaling regime. In the process, we unveil a striking connection of these exact distributions, classically associated with the Jacobi polynomials, with the simpler Legendre polynomials. This new connection allows us to firstly give an explicit proof that shows that the extreme eigenvalue distributions of $\mathbf{W}$ can be expressed in terms of $\alpha_1$- and $\alpha_2$-dimensional determinants involving Bessel functions, without resorting to study correlation kernels. The proof is a direct one, which takes $n$ large in our new exact formulas, and can be viewed as the ``double Wishart analogue'' of a similar proof provided for the LUE in \cite{Forrester1994alpha_det}, but it now boils down to manipulating Legendre polynomials instead of Laguerre polynomials. Secondly, following similar manipulations, we provide finite-$n$ corrections for the extreme eigenvalue distributions of $\mathbf{W}$, giving insights on the universality for the JUE and LUE at the left edge of the spectrum support. To this end, we derive new asymptotic results for the Legendre and associated Legendre polynomials, which are non-standard and may be of independent interest.

  
  

\subsection{Basic Definitions} \label{Section1}

The exact results involve Jacobi, Legendre and associated Legendre polynomials. These are defined as follows. The Jacobi polynomial of degree $l$, and parameters $\alpha$ and $\beta$, admits \cite[eq.~(8.960.1)]{Gradsteyn}
\begin{equation}
  P_l^{(\alpha,\beta)}(x)=\frac{1}{2^l}\sum_{k=0}^l{l+\alpha\choose k}{l+\beta\choose l-k}(x-1)^{l-k}(x+1)^{l}.
\end{equation}
Jacobi polynomials are orthogonal with respect to the weight $w(x)=(1-x)^\alpha(1+x)^{\beta}$ in the interval $[-1,1]$, i.e. \cite[eq.~(22.2.1)]{Abramowitz}
\begin{equation}
\int_{-1}^1(1-x)^\alpha(1+x)^{\beta}{P}^{(\alpha,\beta)}_k(x){P}^{(\alpha,\beta)}_l(x)dx=\frac{2^{\alpha+\beta+1}}{2l+\alpha+\beta+1}\frac{(l+\alpha)!(l+\beta)!}{l!(l+\alpha+\beta)!}\delta_{kl}
\end{equation}
where $\delta_{kl}$ denotes the Kronecker delta function, which equals $1$ if $l = k$ and 0 otherwise. 

The Legendre polynomial of degree $l$ admits \cite[eq.~(8.910.2)]{Gradsteyn}
\begin{equation}
  P_l(x)=\sum_{k=0}^{l}{l\choose k}{l+k\choose k}\left(\frac{x-1}{2}\right)^k.
\end{equation}
Legendre polynomials are particular cases of Jacobi polynomials when $\alpha=\beta=0$. They are then orthogonal with respect to the weight $w(x)=1$ in the interval $[-1,1]$. The associated Legendre polynomial of degree $l$ and order $p$ is defined by \cite[eq.~(8.810)]{Gradsteyn}
\begin{equation} \label{eq:associated_legendre_def}
P_l^p(x)=(-1)^p(1-x^2)^{p/2}\frac{d^pP_l(x)}{dx^p}.
\end{equation} 

Our asymptotic results involve the $l$th order modified Bessel function of the first kind, defined by \mbox{\cite[eq.~(8.406.3)]{Gradsteyn}}
\begin{equation} \label{eq:Bessel_def2}
  I_l(z)=\frac{z^l}{2^l}\sum_{k=0}^\infty\frac{z^{2k}}{2^{2k}k!(l+k)!}
\end{equation}
for $z\in\mathbb{C}$.

\subsection{Models} \label{Section_model}

Let $\mathbf{X}_1\in\mathbb{C}^{m_1\times n}$ ($m_1\geq n$) and $\mathbf{X}_2\in\mathbb{C}^{m_2\times n}$ ($m_2\geq n$) be independent complex Gaussian matrices with independent columns that have the same covariance matrix $\mathbf{\Sigma}$. Then, $\mathbf{W}_1=\mathbf{X}_1^\dagger\mathbf{X}_1$ and $\mathbf{W}_2=\mathbf{X}_2^\dagger\mathbf{X}_2$ are $n\times n$ complex Wishart matrices with $m_1$ and $m_2$ degrees of freedom respectively; i.e., $\mathbf{W}_1\sim\mathcal{CW}_n(m_1, \mathbf{\Sigma})$ and $\mathbf{W}_2\sim\mathcal{CW}_n(m_2, \mathbf{\Sigma})$. Define $\alpha_1=m_1-n$, $\alpha_2=m_2-n$. The joint probability density function (JPDF) of the eigenvalues of 
\begin{equation} \label{eq:model_def}
  \mathbf{W}=(\mathbf{W}_1+\mathbf{W}_2)^{-1}\mathbf{W}_1
\end{equation}
is proportional to \cite{Johnstone2008}
\begin{equation} \label{eq:JPDF_Beta}
  \prod_{k=1}^{n}{\phi}_k^{\alpha_1}(1-{\phi}_k)^{\alpha_2}\prod_{i<j}^n({\phi}_i-{\phi}_j)^2
\end{equation}
with $1\geq {\phi}_1>\ldots>{\phi}_n\geq 0$. This JPDF does not depend on $\mathbf{\Sigma}$ since the eigenvalues of $\mathbf{W}$ do not change under the joint transformation $\mathbf{W}_1\mapsto\mathbf{\Sigma}^{-1/2}\mathbf{W}_1\mathbf{\Sigma}^{-1/2}$, $\mathbf{W}_2\mapsto\mathbf{\Sigma}^{-1/2}\mathbf{W}_2\mathbf{\Sigma}^{-1/2}$ \cite{Johnstone2008}. The ensemble of $n\times n$ matrices of the form (\ref{eq:model_def}) is said to have the multivariate complex beta distribution with parameters $\alpha_1$ and $\alpha_2$ \cite{Muirhead}. The JPDF of its eigenvalues is related to that of other well-known ensembles as follows.

The first is the classical JUE, the ensemble of $n\times n$ matrices with eigenvalue JPDF proportional to
\begin{equation} \label{eq:JPDF_JUE}
  \prod_{k=1}^{n}(1+\widetilde{\phi}_k)^{\alpha_1}(1-\widetilde{\phi}_k)^{\alpha_2}\prod_{i<j}^n(\widetilde{\phi}_i-\widetilde{\phi}_j)^2
\end{equation}
with $1\geq \widetilde{\phi}_1>\ldots>\widetilde{\phi}_n\geq -1$. From (\ref{eq:JPDF_Beta}), one obtains (\ref{eq:JPDF_JUE}) by performing the transformation $\phi_k=(1+\widetilde{\phi}_k)/2$, $k=1,\ldots,n$ \cite{Johnstone2008}. The second is the set of random matrices $\mathbf{F}=\mathbf{W}_1\mathbf{W}_2^{-1}$, commonly referred to as the complex F model \cite{Muirhead}. The JPDF of the eigenvalues  $\infty > \hat{\phi}_1>\ldots>\hat{\phi}_n\geq 0$ of $\mathbf{F}$ is obtained from (\ref{eq:JPDF_Beta}) by performing the transformation $\phi_k=\hat\phi_k/(1+\hat\phi_k)$, $k=1,\ldots,n$ \cite{Chen2012}. It is said that $\mathbf{F}$ and $\mathbf{W}$ are ``matrix analogue'' \cite{Johnstone2008}. 

In this work, we first study the extreme eigenvalue distributions of $\mathbf{W}$, providing new exact determinant expressions which are then leveraged to present asymptotic results and finite-$n$ corrections under hard-edge scaling, i.e., for $n\rightarrow\infty$ with $\alpha_1$ and $\alpha_2$ fixed. By simply applying the corresponding transformations aforementioned, our results for $\mathbf{W}$ can immediately be rephrased for the JUE and F models. 


Before presenting our main results, we make note of the following: 

\begin{remark} \label{remark1}
 One sees that $1-\phi_n$ is the largest eigenvalue of  $\mathbf{I}-(\mathbf{W}_1+\mathbf{W}_2)^{-1}\mathbf{W}_1=(\mathbf{W}_1+\mathbf{W}_2)^{-1}\mathbf{W}_2$. Hence, from the smallest eigenvalue distribution of $\mathbf{W}$, we can deduce that of the largest eigenvalue by simply applying the transformation $\xi\mapsto 1-\xi$ and interchanging $\alpha_1$ with $\alpha_2$ \cite{Muirhead}. 
\end{remark} 
 
 \subsection{Exact extreme eigenvalue distributions of $\mathbf{W}$}
\begin{theorem}
\label{Theorem_min_central}
The cumulative distribution function of the smallest eigenvalue $\phi_n$ of $\mathbf{W}$ admits 
\begin{equation}
\label{eq:smallest_central}
F_{\phi_n}(\xi)=g_{\alpha_1,\alpha_2}(\xi)
\end{equation}
and that of the largest eigenvalue $\phi_1$ admits
\begin{equation}
\label{eq:largest_central}
F_{\phi_1}(\xi)= 1-g_{\alpha_2,\alpha_1}(1-\xi) 
\end{equation}
where
\begin{equation}
\label{eq:H}
g_{\alpha,\beta}(\xi)=1-(1-\xi)^{n^2+n\alpha+n\beta+\alpha\beta}\frac{\det\left[\mathbf{E}_{\alpha}\left(-\frac{1+\xi}{1-\xi}\right)\quad \mathbf{E}_{\beta}(1)\right]}{\det\left[\mathbf{E}_{\alpha}(-1)\quad \mathbf{E}_{\beta}(1)\right]}
\end{equation}
with $\mathbf{E}_{\gamma}(y)$ the $(\alpha+\beta)\times \gamma$ matrix with entries
\begin{equation} \label{eq:E}
[\mathbf{E}_{\gamma}(y)]_{ij}=\frac{d^{j-1}P_{n+i-1}\left(y\right)}{dy^{j-1}}.
\end{equation}
\end{theorem}

The entries in (\ref{eq:E}) admit the explicit representation 
\begin{equation}
\label{eq:g}
[\mathbf{E}_{\gamma}(y)]_{ij}=\begin{cases}
(-1)^{j-1} \left(1-y^2\right)^{-\frac{j-1}{2}}P^{j-1}_{n+i-1}(y), & y \neq 1, y\neq -1\\
2^{1-j}(n+i-j+1)_{2j-2}/(j-1)!, & y=1\\
(-1)^{n+i+j}[\mathbf{E}_{\gamma}(1)]_{ij}, & y=-1.\\
\end{cases}
\end{equation}
Theorem \ref{Theorem_min_central} reveals a tight connection between the distributions of the extreme eigenvalues of $\mathbf{W}$ and Legendre polynomials, which are simple particular cases of the Jacobi and Gegenbauer polynomials \cite{SzegoBook}. The derived exact expressions involve $(\alpha_1+\alpha_2)$-dimensional determinants, whose entries are given exclusively in terms of derivatives of these Legendre polynomials. This has some interesting implications. First, the dimensionality of the determinants in Theorem \ref{Theorem_min_central} does not explode when $n$ grows large, if $\alpha_1$ and $\alpha_2$ are kept fixed. This allows for efficient computation of the extreme eigenvalue distributions in such cases. Moreover, the simple structure of the block matrices inside the determinant in (\ref{eq:E}) is analogous to a determinant representation derived previously for the smallest eigenvalue distribution of the LUE \cite{Forrester1994alpha_det}, which is said to be of \emph{Wronskian-type} (in that case, the successive derivatives were with respect to Laguerre polynomials).  
Due to this analogy, despite the derivation being more challenging, we will show that we can employ similar manipulations to  those presented in \cite{Forrester1994alpha_det} to study the large-$n$ behavior of the extreme eigenvalue distributions.

The results of Theorem \ref{Theorem_min_central} reduce to simplified forms when either $\alpha_1=0$ or $\alpha_2=0$.

\begin{corollary} \label{corollary_2}
  When $\alpha_1=0$,
\begin{equation}
\label{eq:smallest_simple}
F_{\phi_n}(\xi)=1-(1-\xi)^{n^2+n\alpha_2} \; ,
\end{equation}
while when $\alpha_2=0$, 
  \begin{equation}
\label{eq:smallest_simple2}
F_{\phi_n}(\xi)=1- K_{\alpha_1} (1-\xi)^{n^2+n\alpha_1}\det\left(\mathbf{E}_{\alpha_1}\left(\frac{1+\xi}{1-\xi}\right)\right)
\end{equation}
where $K_\alpha = \prod_{k=0}^{\alpha-1}\frac{2^{k-1}}{(2n+2\alpha-2k)_k}$, with $K_0=1$.

Similarly, when $\alpha_2=0$,
\begin{equation} \label{eq:largest_simple}
F_{\phi_1}(\xi)=\xi^{n^2+n\alpha_1} \; ,
\end{equation}
while when $\alpha_1=0$,
\begin{equation} \label{eq:largest_simple2}
F_{\phi_1}(\xi)= K_{\alpha_2} \,  \xi^{n^2+n\alpha_2}  \det\left(\mathbf{E}_{\alpha_2}\left(\frac{2}{\xi}-1\right)\right) .
\end{equation}
\end{corollary}
 
It also turns out that by manipulating known extreme eigenvalue distribution results which were expressed in terms of a Gauss hypergeometric function of a matrix argument (see Table \ref{table_1}), one can also obtain an equivalent expression involving a smaller size determinant, albeit with more complicated entries.  Specifically, the expression for the distribution of the smallest eigenvalue involves an $\alpha_1$-dimensional determinant, while that for the largest eigenvalue involves an $\alpha_2$-dimensional determinant; in both cases the entries involve relatively complicated linear combinations of derivatives of Jacobi polynomials.  The result is as follows:

{
\begin{proposition} \label{Theorem_min_central_Jacobi}
 The cumulative distribution function of $\phi_n$ also admits
\begin{equation}
\label{eq:smallest_central_Forrester}
F_{\phi_n}(\xi)=h_{\alpha_1,\alpha_2}(\xi)
\end{equation}
and that of $\phi_1$ also admits
\begin{equation}
\label{eq:largest_central_Forrester}
F_{\phi_1}(\xi)=1-h_{\alpha_2,\alpha_1}(1-\xi)
\end{equation}
where  
\begin{equation}
h_{\alpha,\beta}(\xi)=1-\prod_{k=1}^{\alpha}\frac{(n+k-1)!(\alpha-k)!}{(\alpha+n-1)!(k-1)!}(1-\xi)^{n^2+n\alpha+n\beta}\det(\mathbf{G})
  \end{equation}
  with $\mathbf{G}$ the $\alpha\times\alpha$ matrix with entries
  \begin{equation} \label{eq:entries_G}
    \begin{split}
    [\mathbf{G}]_{ij}=\ &\sum_{k=0}^{\alpha-j}{\alpha-j\choose k}(-2)^k(j-i+k+1)_{\alpha-j-k}\left(-\frac{\xi}{1-\xi}\right)^{j-i+k}\frac{d^k}{dy^k}P_{n+i-1}^{(\alpha-i,\beta-i+1)}\left(y\right)\Big\vert_{y=\frac{1+\xi}{1-\xi}}.
      \end{split}
      \end{equation}
            \end{proposition}
            
Since the $k$th derivative of a Jacobi polynomial can be expressed in terms of another Jacobi polynomial \cite[eq.~(8.961.4)]{Gradsteyn}, the entries in (\ref{eq:entries_G}) admit the explicit representation 
\begin{equation}
\begin{split}
[\mathbf{G}]_{ij}=\ &\sum_{k=0}^{\alpha-j}{\alpha-j\choose k}(-1)^k(j-i+k+1)_{\alpha-j-k}(n+\alpha+\beta-i+1)_k\\
&\times\left(-\frac{\xi}{1-\xi}\right)^{j-i+k}P_{n+i-k-1}^{(\alpha+k-i,\beta+k-i+1)}\left(\frac{1+\xi}{1-\xi}\right).
\end{split}
\end{equation}

It is noteworthy that the simplified special cases (\ref{eq:smallest_simple}) and (\ref{eq:largest_simple}) are also easily recoverable from this proposition; however directly recovering the simplified forms (\ref{eq:smallest_simple2}) and (\ref{eq:largest_simple2}) does not appear straightforward.  Moreover, due to its simplified structure and dependence on Legendre polynomials (which, recall, are simplified cases of Jacobi polynomials), the results in Theorem \ref{Theorem_min_central} are more amenable to direct asymptotic analysis than those given in Proposition \ref{Theorem_min_central_Jacobi}.  We now pursue such asymptotic analysis.


\subsection{Asymptotic extreme eigenvalue distributions of $\mathbf{W}$}

We consider the hard-edge scaling limit, for which $n$ grows large, with $\alpha_1$ and $\alpha_2$ fixed. Results under this scaling have been considered previously, where the eigenvalue correlation kernel of the JUE (with appropriate centering and scaling) has been shown to coincide with that of the LUE in this asymptotic limit \cite{Nagao1993}. This implies that the extreme eigenvalue distributions of the JUE should coincide with the distribution of the smallest eigenvalue of the LUE, which has been shown to admit a remarkably simple form involving a finite-dimensional determinant whose entries are Bessel functions \cite{Forrester1994alpha_det}. Using this correspondence, along with the simple mapping between the eigenvalues of ${\bf W}$ and the JUE, given in Section \ref{Section_model}, an analogous determinant expression should be obtained for the extreme eigenvalue distributions of ${\bf W}$.

Here we provide a direct proof of this result, without resorting to a study of correlation kernels etc., by simply taking $n$ large in our exact formulas for the extreme eigenvalues of $\mathbf{W}$.  This is accomplished by deriving asymptotic expansions of Legendre polynomials that are non-standard, and may be of independent interest. In principle, this direct approach is the ``double Wishart analogue'' (or ``JUE analogue'') of a similar direct proof provided for the LUE in \cite{Forrester1994alpha_det}, which exploited asymptotic properties of Laguerre polynomials.  Our derivation, while being based on  elementary operations, also enables explicit computation of the large (but finite) $n$ correction terms to the asymptotic distribution. We present this for some particular cases of $\alpha_1$ and $\alpha_2$.  

To guide our asymptotic analysis, it is insightful to first study the scaling of the mean and standard deviation of the smallest eigenvalue, using the simple representation in (\ref{eq:smallest_simple}).  Specifically, explicit computation of the mean yields
\begin{equation}
  \begin{split}
\mathbb{E}[\phi_n]=&\frac{1}{n^2+n\alpha_2+1}\\
=&\frac{1}{n^2}+o\left(\frac{1}{n^2}\right)
\end{split}
\end{equation}
while, for the standard deviation,
\begin{equation}
\label{eq:scaling_corollary_smallest}
\begin{split}
\sigma_{\phi_n}=&\sqrt{\frac{n(n+\alpha_2)}{(1+n(n+\alpha_2))^2(2+n(n+\alpha_2))}}\\
=&\frac{1}{n^2}+o\left(\frac{1}{n^2}\right).
\end{split}
\end{equation}
It is therefore natural to scale $\phi_n$ by $n^2$ to study its asymptotic distribution (see also \cite{Nagao1993}). Recall also that the asymptotic distribution of the largest eigenvalue can be deduced from that of the smallest one, as indicated in \mbox{Remark \ref{remark1}}. With this in mind, defining 
\begin{equation} \label{eq:F_inf}
    F_\infty^{(\alpha)}(x)=1-e^{-x}\det\left[I_{i-j}(\sqrt{4x})\right]_{i,j=1,\ldots,\alpha} ,
\end{equation}
we arrive at the following:

\begin{theorem}
 \label{Theorem_asymptoticcalpha1xalpha1}
 For fixed $\alpha_1$ and $\alpha_2$,
   \begin{equation} \label{eq:asympt_min}
  \lim_{n\rightarrow\infty}F_{n^2\phi_n}\left(x\right)=\lim_{n\rightarrow\infty}F_{\phi_n}\left(\frac{x}{n^2}\right) = F_\infty^{(\alpha_1)}(x)
  \end{equation}
  and
  \begin{equation} \label{eq:asympt_max}
  \lim_{n\rightarrow\infty}F_{n^2(1-\phi_1)}(x)=1-\lim_{n\rightarrow\infty}F_{\phi_1}\left(1-\frac{x}{n^2}\right) = F_\infty^{(\alpha_2)}(x)
    \end{equation}
    for $x\geq 0$.
\end{theorem}
   
}


Contrasting this result with Theorem \ref{Theorem_min_central}, where the exact extreme eigenvalue distributions of $\mathbf{W}$ were given in terms of $(\alpha_1+\alpha_2)$-dimensional determinants, the asymptotic distributions in Theorem \ref{Theorem_asymptoticcalpha1xalpha1} 
involve $\alpha_1$- or $\alpha_2$-dimensional determinants, as in Proposition \ref{Theorem_min_central_Jacobi}. However, contrary to Proposition \ref{Theorem_min_central_Jacobi}, other than in defining the determinant size, there is no further dependence on either $\alpha_1$ or $\alpha_2$ in the asymptotic expression.  Hence, for both the largest and smallest eigenvalue distributions, the dependence on one of the alphas is fully washed out when taking hard-edge asymptotics, while the dependence on the other is simply to determine the dimensionality of the matrix determinant.  


If one now considers the JUE, for which $\tilde{\phi}_k = 2 \phi_k -1$ (see Section \ref{Section_model}), we easily establish that
\begin{align} \label{eq:BetaJUETrans}
& F_{n^2 ( \tilde{\phi}_n + 1 )}(x) = F_{n^2 \phi_n}(x/2) \; , \nonumber \\
& F_{n^2 ( 1 - \tilde{\phi}_1 )}(x) = F_{n^2 ( 1 - \phi_n ) }(x/2) \; ,
\end{align}
and therefore
\begin{align}
&\lim_{n\rightarrow\infty} F_{n^2 ( \tilde{\phi}_n + 1 )}(x) = F_\infty^{(\alpha_1)}(x / 2) , \nonumber \\
&\lim_{n\rightarrow\infty} F_{n^2 ( 1 - \tilde{\phi}_1 )}(x) = F_\infty^{(\alpha_2)}(x / 2) \; .
\end{align}

These asymptotic distributions coincide precisely with the smallest eigenvalue distribution of the LUE under similar hard-edge scaling (with suitable parameterization of $\alpha_1$ and $\alpha_2$), suggesting a form of ``universality'' under the hard-edge scaling limit. While this is aligned with previous results relating the hard-edge scaling of the JUE and LUE \cite{Nagao1993}, an open question is whether such correspondence persists when considering first-order correction terms to the asymptotic distribution?  Recently, these correction terms were computed explicitly for the LUE \cite{Edelman2014beyond}, though for the JUE (or the double Wishart model) we are unaware of any corresponding results. The explicit exact eigenvalue distribution in Theorem \ref{Theorem_min_central}  lends itself to this analysis, at least for specific values of $\alpha_1$ or $\alpha_2$, as we present below.  A generalized formula for arbitrary $\alpha_1$, $\alpha_2$ may also be possible, although we have been unable to establish a generalized proof at this point.    

We first recall that the density corresponding to the asymptotic distribution (\ref{eq:F_inf}) admits \cite{Forrester1994alpha_det}
\begin{equation} \label{eq:density_Forrester}
    f_\infty^{(\alpha)}(x)= \frac{d}{d x} F_\infty^{(\alpha)}(x) = e^{-x}\det\left[I_{2+i-j}(\sqrt{4x})\right]_{i,j=1,\ldots,\alpha} .
\end{equation}
Our main result is the following: 

\begin{proposition} \label{prop_correction1}
  For $\alpha_1=0,1$ and arbitrary (but fixed) $\alpha_2$,
   \begin{equation} \label{eq:suggestion}
  F_{n^2\phi_n}\left(x\right)
=F_{\infty}^{(\alpha_1)}(x)+\frac{\alpha_1+\alpha_2}{n}x f_\infty^{(\alpha_1)}(x) +\mathcal{O}\left(\frac{1}{n^2}\right)\; .
  \end{equation}
  This also holds for $\alpha_1=2$ and $\alpha_2=0,1,2$.
  
  Similarly, for arbitrary $\alpha_1$ and $\alpha_2=0,1$,
  \begin{equation} \label{eq:suggestion2}
  F_{n^2(1-\phi_1)}\left(x\right)
=F_{\infty}^{(\alpha_2)}(x)+\frac{\alpha_1+\alpha_2}{n}x f_\infty^{(\alpha_2)}(x) +\mathcal{O}\left(\frac{1}{n^2}\right)\; ,
  \end{equation}
  which also holds for $\alpha_1=0,1,2$ and $\alpha_2=2$. 
\end{proposition}

Proposition \ref{prop_correction1} reveals that, for the cases of $\alpha_1$,$\alpha_2$ considered, the first-order correction of the extreme eigenvalue distributions of $\mathbf{W}$ is proportional to the density (\ref{eq:density_Forrester}) which, similar to the asymptotic distributions of Theorem \ref{Theorem_asymptoticcalpha1xalpha1}, is given as an $\alpha_1$- or $\alpha_2$-dimensional determinant.
Interestingly, the alpha parameter which was washed out in those asymptotic expressions appears when considering finite-$n$ corrections, as a scaling factor in the first-order correction term of (\ref{eq:suggestion})-(\ref{eq:suggestion2}).
From the equivalence (\ref{eq:BetaJUETrans}), Proposition \ref{prop_correction1} can be immediately rephrased for the JUE. Focusing in particular on the smallest eigenvalue, and for the cases of $\alpha_1$,$\alpha_2$ considered in the proposition,  
\begin{equation} \label{eq:correction_JUE}
    F_{n^2(1+\tilde{\phi}_n)}(x)=F_{\infty}^{(\alpha_1)}\left(\frac{x}{2}\right)+\frac{\alpha_1+\alpha_2}{2n}x f_\infty^{(\alpha_1)}\left(\frac{x}{2}\right) +\mathcal{O}\left(\frac{1}{n^2}\right),
\end{equation}
which bears a strong analogy with a recent corresponding result for the  LUE, conjectured in \cite{Edelman2014beyond} and proved in \cite{Perret2016, Bornemann2016}; specifically, for the LUE with fixed parameter $\alpha$, the distribution of the smallest eigenvalue for large (but finite) $n$ is given by \cite[Theorem 4.2]{Edelman2014beyond} 
\begin{equation} \label{eq:correction_LUE}
     F_{\rm LUE}(x)=F_{\infty}^{(\alpha)}\left(\frac{x}{2}\right)+\frac{\alpha}{2n}x f_\infty^{(\alpha)}\left(\frac{x}{2}\right) +\mathcal{O}\left(\frac{1}{n^2}\right),
\end{equation}
which coincides with that of the JUE when $\alpha_1=\alpha$ and $\alpha_2=0$. Therefore, Proposition \ref{prop_correction1} shows that the correspondence under the hard-edge scaling between the extreme eigenvalue distributions of the JUE and LUE still holds for finite-$n$ corrections to first order, at least for the specific values of $\alpha_1, \alpha_2$ considered.

A natural question is whether the result of Proposition \ref{prop_correction1} and, therefore, the suggested universality of the first-order corrections under hard-edge scaling is still valid for arbitrary $\alpha_1$ and $\alpha_2$. The proof of the general case is particularly challenging, due to the overwhelming number of terms that appear in the iterative procedure to reduce the dimensions of the involved determinants (see Section \ref{sec:proofProp2}). Although we have not been able to establish such proof, in the following we present some numerical results which both validates Proposition \ref{prop_correction1}, and checks numerically whether the stated first-order corrections may continue to hold beyond the cases of $\alpha_1$, $\alpha_2$ considered in Proposition \ref{prop_correction1}.

\begin{figure}
\centering
\begin{subfigure}[t]{0.48\textwidth} \centering
\includegraphics[width=\linewidth]{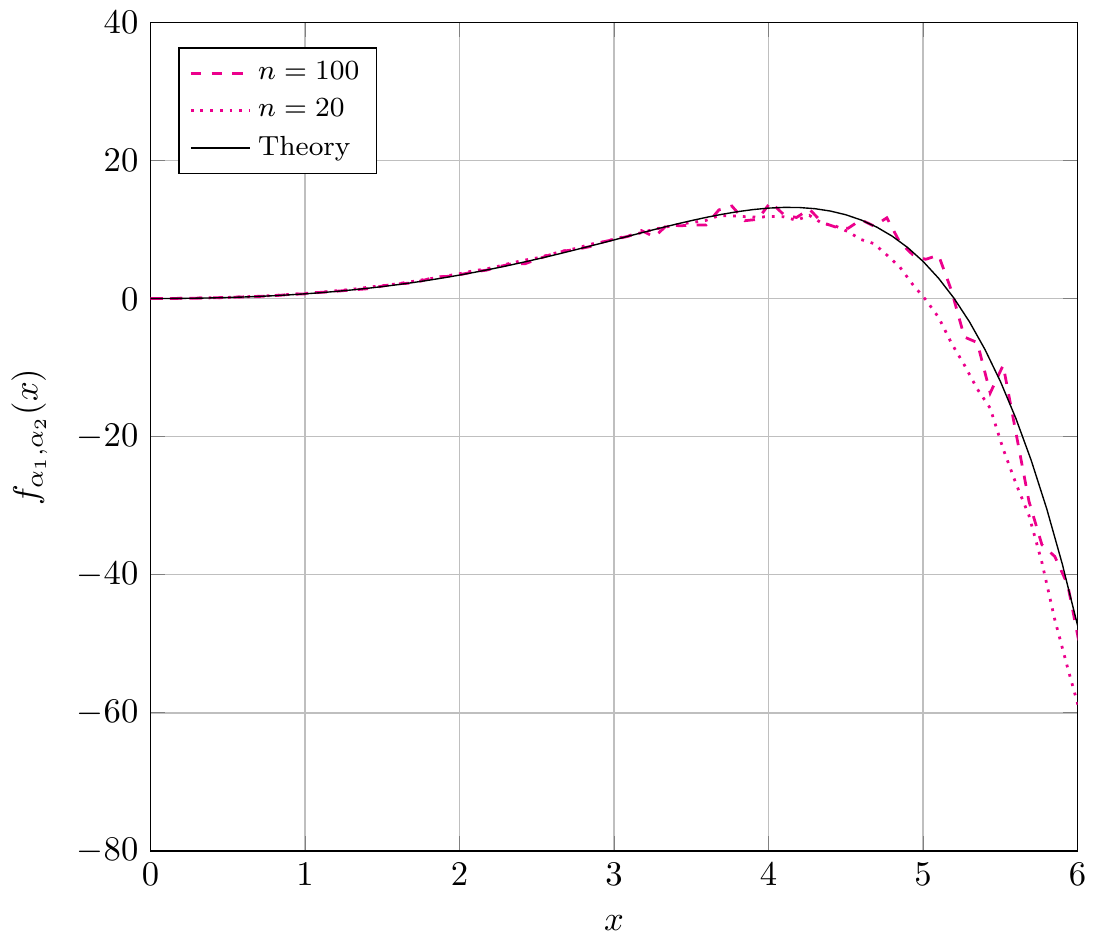}
\caption{$\alpha_1=2$ and $\alpha_2=0$} \label{fig:alpha1_2_alpha2_0}
\end{subfigure}
\begin{subfigure}[t]{0.48\textwidth} \centering
\includegraphics[width=\linewidth]{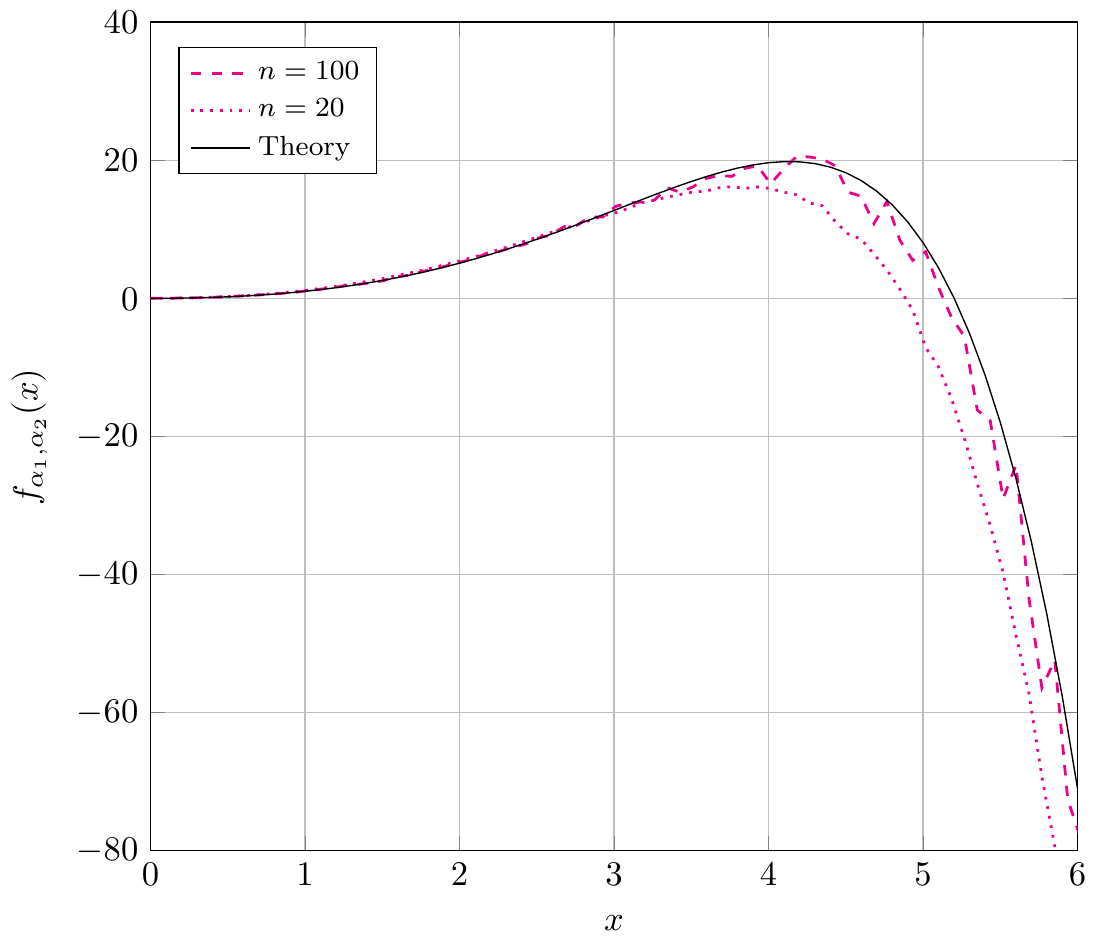}
\caption{$\alpha_1=2$ and $\alpha_2=1$} \label{fig:alpha1_2_alpha2_1}
\end{subfigure}
\begin{subfigure}[t]{0.48\textwidth} \centering
\includegraphics[width=\linewidth]{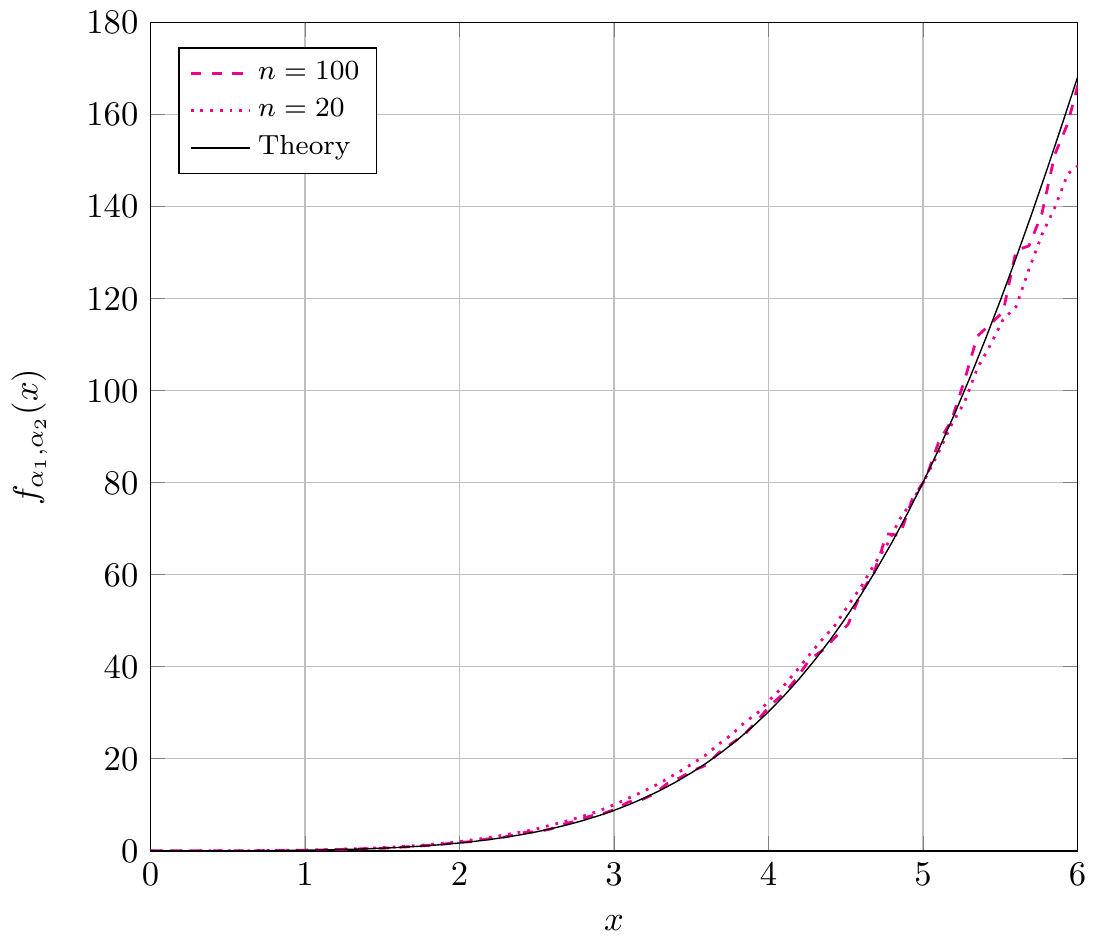}
\caption{$\alpha_1=3$ and $\alpha_2=0$} \label{fig:alpha1_3_alpha2_0}
\end{subfigure}
\begin{subfigure}[t]{0.48\textwidth} \centering
\includegraphics[width=\linewidth]{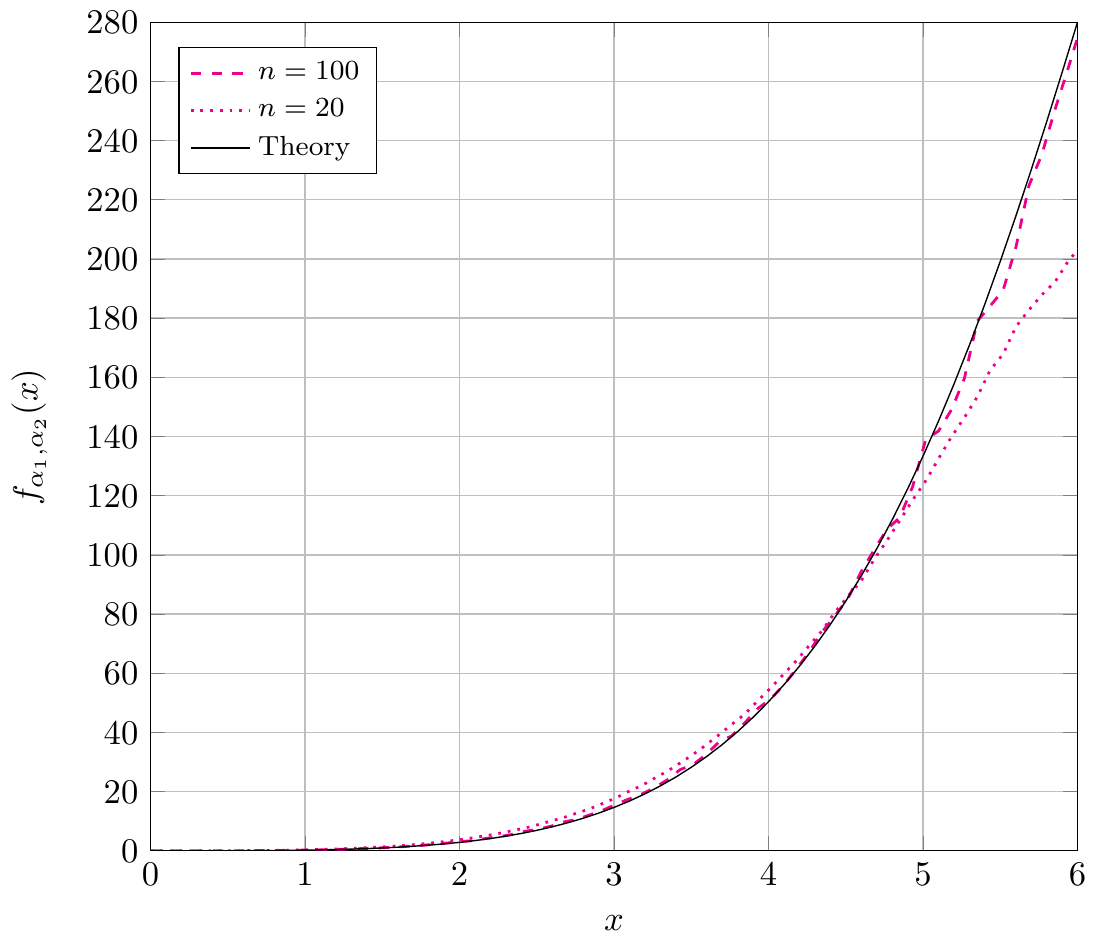}
\caption{$\alpha_1=3$ and $\alpha_2=2$}  \label{fig:alpha1_3_alpha2_2}
\end{subfigure}
\caption{Scaled first-order correction (\ref{eq:scaled_correction}) for different values of $\alpha_1$ and $\alpha_2$. Monte Carlo simulations are plotted with the theoretical zeroth-order term subtracted and the result is multiplied by $ne^{x}$.} 
\end{figure}

We first computed the empirical density of the smallest eigenvalue of $\mathbf{W}$ for the cases $\alpha_1=2$ and $\alpha_2=0,1$.
This was computed from 50 million realizations of $\mathbf{W}$ for $n=20$ and $n=100$; the simulations took $11$ hours for $n=20$ and $37$ hours for $n=100$ on a $12$-core computer. In Figs. \ref{fig:alpha1_2_alpha2_0} and \ref{fig:alpha1_2_alpha2_1} we show the empirical correction, computed as the difference between the empirical density and the theoretical asymptotic density $f_{\infty}^{(\alpha_1)}(x) = \frac{d}{dx} F_{\infty}^{(\alpha_1)}(x)$, scaled\footnote{We find it convenient to scale the correction term by $n e^{x}$, as opposed to simply by $n$, to cancel the exponential factor that appears from the derivative in (\ref{eq:scaled_correction}), which would otherwise dominate the behavior of the correction term, rendering numerical validations visually less clear.} by $n e^x$, along with the theoretical first-order correction to the asymptotic density, obtained from (\ref{eq:suggestion}) in Proposition \ref{prop_correction1} (and correspondingly scaled) as
\begin{equation} \label{eq:scaled_correction}
    f_{\alpha_1,\alpha_2}(x)=e^{x}(\alpha_1+\alpha_2)\frac{d}{dx}\left[xf_\infty^{(\alpha_1)}(x)\right],
\end{equation}
which gives, for $\alpha_1=2$ and arbitrary $\alpha_2$,
\begin{equation}
\begin{split}
    f_{2,\alpha_2}(x)=\ &(2+\alpha_2)(1-x)\left(I_0(\sqrt{4x})^2-\left(1+\frac{1}{x}\right)I_1(\sqrt{4x})^2\right)\\
    &+(2+\alpha_2)x\left(I_1(\sqrt{4x})^2\left(\frac{1}{x}+\frac{2}{x^2}\right)-\frac{2}{x\sqrt{x}}I_0(\sqrt{4x})I_1(\sqrt{4x})\right).
    \end{split}
\end{equation}
As expected, the simulated correction approaches the theoretical first-order correction as $n$ increases, since the contribution of higher-order terms in the simulated correction decreases. The agreement between the simulated and theoretically-predicted correction is already evident at $n=100$. 

To further evaluate whether the theoretical first-order correction holds beyond the cases of \mbox{Proposition \ref{prop_correction1}}, we again computed the empirical density and compared the empirical correction with the theoretical one, just as in Figs. \ref{fig:alpha1_2_alpha2_0} and \ref{fig:alpha1_2_alpha2_1}, but now for $\alpha_1=3$ and $\alpha_2=0,2$. The results are presented in Figs. \ref{fig:alpha1_3_alpha2_0} and \ref{fig:alpha1_3_alpha2_2}, which again show an excellent agreement between simulated and theoretically-predicted corrections, suggesting that Proposition \ref{prop_correction1} may hold in general. This is formally conjectured as follows:





\begin{conjecture} \label{conjecture}
For arbitrary $\alpha_1$ and $\alpha_2$,
\begin{align} \label{eq:conjecture}
  F_{n^2\phi_n}\left(x\right)
&=F_{\infty}^{(\alpha_1)}(x)+\frac{\alpha_1+\alpha_2}{n}x f_\infty^{(\alpha_1)}(x) +\mathcal{O}\left(\frac{1}{n^2}\right)\; , \\
F_{n^2(1-\phi_1)}\left(x\right)
&=F_{\infty}^{(\alpha_2)}(x)+\frac{\alpha_1+\alpha_2}{n}x f_\infty^{(\alpha_2)}(x) +\mathcal{O}\left(\frac{1}{n^2}\right)\; .
  \end{align}
\end{conjecture}
With this, we equivalently conjecture that the first-order corrections to the asymptotic distribution for the extreme eigenvalues of the JUE are indeed equivalent to those for the smallest eigenvalue of the LUE in general, upon suitable JUE-LUE parametrization; recall that the LUE is parametrized by a single alpha, so that for the equivalence with the JUE to hold, we must either have $\alpha_1=0$ or $\alpha_2=0$ when respectively considering the largest or the smallest eigenvalue of the JUE. When both $\alpha_1$ and $\alpha_2$ are non-zero, the suggested universality of the first-order corrections under hard-edge scaling does not persist, due to the scaling factor $(\alpha_1+\alpha_2)$ in the correction terms given in Conjecture \ref{conjecture}. 

A further interesting question is whether the correspondence between the LUE and JUE, and the suggested universality under the hard-edge scaling, still hold for second-order (or higher-order) correction terms, upon suitable JUE-LUE parametrization. An answer to this question requires substantial further analysis, and remains an interesting topic for future investigation.

\section{Legendre Polynomials and Bessel Functions} \label{section:legendre}

\label{Section_preliminaries_proof}

Our analysis relies heavily on properties of Legendre polynomials and their asymptotic connection to Bessel functions. In this section, we summarize the properties needed for the proofs of Theorem \ref{Theorem_min_central}, Theorem \ref{Theorem_asymptoticcalpha1xalpha1} and Proposition \ref{prop_correction1}.

\subsection{Additional Definitions}

The Legendre polynomial $P_n(x)$ is alternatively defined by the Rodrigues' formula \cite[eq.~(22.11.5)]{Abramowitz}
  \begin{equation} \label{eq:derivative_Legendre_binomial}
    P_n(x)=\frac{1}{2^n n!}\frac{d^n}{dx^n}(x^2-1)^n
  \end{equation}
  with $P_n(1)=1$, where it is clear that \cite[eq.~(22.4.6)]{Abramowitz}
  \begin{equation} \label{eq:legendre_minus_argument}
    P_n(-x)=(-1)^nP_n(x).
  \end{equation}
  
  The associated Legendre polynomial $P_n^{-m}(x)$ is defined by the Rodrigues' formula
  \begin{equation}\label{eq:Rodrigues_formula_assoc}
    P_n^{-m}(x)=\frac{(-1)^m}{2^n n!}(1-x^2)^{-m/2}\frac{d^{n-m}}{dx^{n-m}}(x^2-1)^n
  \end{equation}
  where $n\geq m$.

The shifted Legendre polynomial of degree $n$ is defined by
\begin{equation} \label{eq:def_shifted}
  \widetilde{P}_n(x)=P_n(2x-1).
\end{equation}
Shifted Legendre polynomials are orthogonal with respect to $1$ in the interval $[0,1]$, i.e., \cite[eq.~(22.2.11)]{Abramowitz}
\begin{equation}
\int_0^1\widetilde{P}_l(x)\widetilde{P}_n(x)dx=\frac{1}{2n+1}\delta_{ln}.
\end{equation}

\subsection{Identities}

\begin{lemma} \label{property2} 
  For $m\geq -(n+1)$ and $n\geq 0$
\begin{equation} \label{eq:Property2_rewrite}
  (n-m+1)\frac{d^{n+m+1}}{dx^{n+m+1}}\left[(x^2-1)^{n+1}\right]=x\frac{d^{n+m+2}}{dx^{n+m+2}}\left[(x^2-1)^{n+1}\right]-2(n+1)\frac{d^{n+m+1}}{dx^{n+m+1}}\left[(x^2-1)^{n}\right].
\end{equation}

\end{lemma}
\begin{proofof}
  To prove such result, we manipulate recurrence properties of associated Legendre polynomials.
  We start with \cite[eq.~(8.731.1)]{Gradsteyn}
  \begin{equation} \label{eq:proofLemma1_1}
    (x^2-1)\frac{d}{dx}P^m_{n}(x)=(n-m+1)P_{n+1}^m(x)-(n+1)xP_{n}^{m}(x)
  \end{equation}
  and \cite[eq.~(8.731.1(1))]{Gradsteyn}
  \begin{equation} \label{eq:proofLemma1_2}
    (x^2-1)\frac{d}{dx}P^m_{n}(x)=n x P_{n}^m(x)-(n+m)P_{n-1}^{m}(x).
  \end{equation}
  Applying the Rodrigues' formula (\ref{eq:Rodrigues_formula_assoc}) and the chain rule, we rewrite (\ref{eq:proofLemma1_1}) and (\ref{eq:proofLemma1_2}) as
\begin{equation} \label{eq:recurr1}
    \frac{(x^2-1)}{2^n n!}\frac{d^{n+m+1} }{dx^{n+m+1}}\left[(x^2-1)^n\right]=\frac{(n-m+1)}{2^{n+1}(n+1)!}\frac{d^{n+m+1} }{dx^{n+m+1}}\left[(x^2-1)^{n+1}\right]-\frac{(n+m+1)x}{2^{n}n!}\frac{d^{n+m} }{dx^{n+m}} \left[(x^2-1)^{n}\right]
  \end{equation}
  and 
\begin{equation} \label{eq:recurr2}
  \frac{(x^2-1)}{2^n n!}\frac{d^{n+m+1} }{dx^{n+m+1}}\left[(x^2-1)^n\right]= \frac{(n-m)x}{2^{n}n!}\frac{d^{n+m} }{dx^{n+m}} \left[(x^2-1)^{n}\right]-\frac{(n+m)}{2^{n-1}(n-1)!}\frac{d^{n+m-1} }{dx^{n+m-1}}\left[(x^2-1)^{n-1}\right]
\end{equation}
respectively. Replacing $n$ with $n+1$ in (\ref{eq:recurr2}), multiplying by $x$ and then subtracting (\ref{eq:recurr1}), we obtain the result after some simplifications.  \qed
\end{proofof}

\begin{corollary} \label{corollary_property2}
  For $m\geq 0$,
  \begin{equation} \label{eq:corollary_property}
  (n-m+1)\frac{d^{m}}{dx^{m}}P_{n+1}(x)=x\frac{d^{m+1}}{dx^{m+1}}P_{n+1}(x)-\frac{d^{m+1}}{dx^{m+1}}P_n(x).
\end{equation}
\end{corollary}
\begin{proofof}
  The proof is straightforward from Lemma \ref{property2} and the Rodrigues' formula (\ref{eq:derivative_Legendre_binomial}).
\end{proofof}

Corollary \ref{corollary_property2} is a special case of Lemma \ref{property2} and will be key to give insight on the proof of Theorem \ref{Theorem_asymptoticcalpha1xalpha1} in Section \ref{proof_3x3}.

\begin{lemma} \label{Lemma_legendre2}
For $-n \leq m \leq n$,
  \begin{equation} \label{eq:Property3}
  \begin{split}
  \frac{1}{2^{n}n!}\frac{d^{n+m}}{dx^{n+m}}\left[(x^2-1)^{n}\right]=&\frac{(n+m)!}{(n-m)!}(1-x^2)^{-m/2}P_{n}^{-m}(x).  
  \end{split}
\end{equation}
\end{lemma}

\begin{proofof}
   This follows from (\ref{eq:Rodrigues_formula_assoc}) and \cite[eq.~(8.752.2)]{Gradsteyn}
\begin{equation} \label{eq:Gradsteyn_assoc}
  P_n^{m}(x)=(-1)^m\frac{(n+m)!}{(n-m)!}P_n^{-m}(x). 
\end{equation}\qed
\end{proofof}
 
 \subsection{Asymptotics}

In \cite[Lemma 4.1.]{Edelman2014beyond}, Edelman et al. provided the following two-term asymptotic expansion for Laguerre polynomials
\begin{equation} \label{eq:asympt_edelman}
    n^{-m}L_{n-m}^{(m)}\left(-\frac{x}{n}\right)=\frac{I_m(2\sqrt{x})}{x^{m/2}}-\frac{1}{2n}\frac{I_{m-2}(2\sqrt{x})}{x^{(m-2)/2}}+\mathcal{O}\left(\frac{1}{n^2}\right),
\end{equation}
which extends the result \mbox{\cite[eq. (3.29)]{Forrester1994alpha_det}}, with $L_l^{(p)}(x)$ the associated Laguerre polynomial of degree $l$ and order $p$. Here, we provide an analogous property for derivatives of Legendre polynomials and associated Legendre polynomials. Our results extend the classical result by Laurent \cite[Section IV]{Laurent1875}
\begin{equation} 
  \lim_{n\rightarrow\infty}P_n\left(\frac{1+z^2/n^2}{1-z^2/n^2}\right)=I_0(2z) \,, \qquad z\in\mathbb{C} .
  \end{equation}

\begin{lemma} \label{Lemma_legendre_correction}
For fixed $m\geq 0$, $c\in\mathbb{Z}$ and $x>0$,
\begin{equation} \label{eq:asympt_derivative_general}
  n^{-2m}\frac{d^m}{dy^m}P_{n+c}(y)\bigg\vert_{y=\frac{1+x/n^2}{1-x/n^2}}=\frac{I_m(\sqrt{4x})}{(4x)^{m/2}}+\frac{1+2c}{2n}\frac{I_{m-1}(\sqrt{4x})}{(4x)^{(m-1)/2}}+\mathcal{O}\left(\frac{1}{n^2}\right).
\end{equation}
For fixed $m,c\in\mathbb{Z}$ and $x\geq 0$,
\begin{equation}
  \begin{split}
  \label{eq:Watson_modif_transf_general}
   n^{-m}P_{n+c}^{m}\left(\frac{1+x/n^2}{1-x/n^2}\right)=(-\iota)^m I_m(\sqrt{4x})+\frac{(-\iota)^{m}}{n}(1+2c)\sqrt{x}I_{m-1}(\sqrt{4x})+\mathcal{O}\left(\frac{1}{n^2}\right).
  \end{split}
\end{equation} 
 where $\iota^2=-1$.
\end{lemma}

\begin{proofof}
  First, we prove (\ref{eq:asympt_derivative_general}) by following the strategy of \cite[Section IV]{Laurent1875}. Let $n> m$.  Using (\ref{eq:derivative_Legendre_binomial}), we rewrite
  \begin{equation}
    \frac{d^m}{dy^m}P_{n+c}(y)=\frac{1}{2^n n!}\frac{d^{n+c+m}}{dy^{n+c+m}}\left[(y-1)^{n+c}(y+1)^{n+c}\right].
  \end{equation}
 Applying Leibnitz formula for the $(n+c+m)$-times differentiation of the product of functions $f(y)=(y+1)^{n+c}$ and $g(y)=(y-1)^{n+c}$, i.e.,
 \begin{equation}
   (fg)^{(n+c+m)}(y)=\sum_{k=m}^{n+c}{n+c+m\choose k}f^{(n+c+m-k)}(y)g^{(k)}(y),
 \end{equation}
 yields 
 \begin{equation}
   \begin{split}
   \frac{d^m}{dy^m}P_{n+c}(y)=\ &\frac{(n+c+m)!}{2^n(n+c-m)!}\left(\frac{1}{m!}(y-1)^{n+c-m}+\frac{(n+c)(n+c-m)}{(m+1)!}(y-1)^{n+c-m-1}(y+1)\right.\\
   &+\left.\frac{(n+c)(n+c-1)(n+c-m)(n+c-m-1)}{2!(m+2)!}(y-1)^{n+c-m-2}(y+1)^2 +\ldots\right),
   \end{split}
 \end{equation}
 or, equivalently, 
 \begin{equation} \label{eq:key_nplusc}
   \begin{split}
   \frac{d^m}{dy^m}P_{n+c}(y)=\ &\frac{(n+c+m)!}{2^{n+c}(n+c-m)!}\left(\frac{1}{m!}+\frac{(n+c)(n+c-m)}{(m+1)!}\frac{y+1}{y-1}\right. \\
   &+\left.\frac{(n+c)(n+c-1)(n+c-m)(n+c-m-1)}{ 2!(m+2)!}\left(\frac{y+1}{y-1}\right)^2 +\ldots\right)(y-1)^{n+c-m}.
   \end{split}
 \end{equation}

 Let 
 \begin{equation} \label{eq:change_variables_lemma3}
   y=-\frac{1+x/n^2}{1-x/n^2};
 \end{equation}
 we have

 \begin{equation}
 \begin{split}
      \frac{d^m}{dy^m}P_{n+c}(y)\Big\vert_{y =- \frac{1+x/n^2}{1-x/n^2}} =&(-1)^{n+c-m}\frac{(n+c+m)!}{2^m(n+c-m)!}\left(\frac{1}{m!}+\frac{(n+c)(n+c-m)}{(m+1)!}\frac{x}{n^2}\right.\\
    &  +\left.\frac{(n+c)(n+c-1)(n+c-m)(n+c-m-1)}{2!(m+2)!}\left(\frac{x}{n^2}\right)^2 +\ldots\right)(1-x/n^2)^{m-n-c}.
 \end{split}
 \end{equation}
 Applying (\ref{eq:legendre_minus_argument}), we obtain
 \begin{equation} \label{eq:derivative_Legendre_expansion}
 \begin{split}
      \frac{d^m}{dy^m}P_{n+c}(y)\Big\vert_{y = \frac{1+x/n^2}{1-x/n^2}} =\ &\frac{(n+c+m)!}{2^m(n+c-m)!}\left(\frac{1}{m!}+\frac{(n+c)(n+c-m)}{(m+1)!}\frac{x}{n^2}\right.\\
    &  +\left.\frac{(n+c)(n+c-1)(n+c-m)(n+c-m-1)}{2!(m+2)!}\left(\frac{x}{n^2}\right)^2 +\ldots\right)(1-x/n^2)^{m-n-c}.
 \end{split}
 \end{equation}
 Now, we are ready to make $n$ large. Since
 \begin{equation} \label{eq:factorial_large}
   \frac{(n+c+m)!}{(n+c-m)!}= n^{2m}+m(1+2c) n^{2m-1}+\mathcal{O}\left(n^{2m-2}\right),
 \end{equation}
 \begin{align}
     &\frac{\left[(n+c)(n+c-1)\ldots(n+c-k)\right]\left[(n+c-m)(n+c-m-1)\ldots(n+c-m-k)\right]}{n^{2(k+1)}} \notag \\
     & \hspace{8cm} =1+\frac{1}{n}(2c-m-k)
     (k+1)
     +\mathcal{O}\left(\frac{1}{n^2}\right)
     \end{align}
 for $k=0,1,2,\ldots$, and
 \begin{equation}
   (1-x/n^2)^{m-n-c}=1+\frac{x}{n}+\mathcal{O}\left(\frac{1}{n^2}\right),
 \end{equation}
we obtain that
 \begin{equation} \label{eq:asympt_derivative2}
 \begin{split}
 n^{-2m}\frac{d^m}{dy^m}P_{n+c}(y)\Big\vert_{y=\frac{1+x/n^2}{1-x/n^2}}&=\frac{1}{2^m}\left(\ldots+\frac{x^{k+1}}{(k+1)!(m+k+1)!}r_k(x)+\ldots\right)
   \end{split}
   \end{equation}
  where we have only presented the $(k+2)$th term of (\ref{eq:derivative_Legendre_expansion}) with
  \begin{equation}
  \begin{split}
      r_k(x)&=\left(1+\frac{m(1+2c)}{n}\right)\left(1+\frac{1}{n}(2c-m-k)(k+1)\right)\left(1+\frac{x}{n}\right)+\mathcal{O}\left(\frac{1}{n^2}\right)\\
   &=\left(1+\frac{1}{n}\left(2c-k)\left(m+k+1\right)+x\right)\right)+\mathcal{O}\left(\frac{1}{n^2}\right).
   \end{split}
  \end{equation}
  Therefore,
  \begin{equation} \label{eq:asympt_derivative2}
 \begin{split}
 n^{-2m}\frac{d^m}{dy^m}P_{n+c}(y)\Big\vert_{y=\frac{1+x/n^2}{1-x/n^2}}=\ &\frac{1}{2^m}\sum_{k=0}^\infty\frac{x^k}{k!(m+k)!}+\frac{2c}{n2^m}\sum_{k=0}^\infty\frac{x^k}{k!(m+k-1)!}\\
 &-\frac{1}{n2^m}\left(\sum_{k=0}^\infty\frac{(k-1)x^k}{k!(m+k-1)!}-x\sum_{k=0}^\infty\frac{x^k}{k!(m+k)!}\right)+\mathcal{O}\left(\frac{1}{n^2}\right)\\
 =\ &\frac{1}{2^m}\sum_{k=0}^\infty\frac{x^k}{k!(m+k)!}+\frac{1+2c}{n2^m}\sum_{k=0}^\infty\frac{x^k}{k!(m+k-1)!}+\mathcal{O}\left(\frac{1}{n^2}\right),
   \end{split}
   \end{equation}
   which leads to the result (\ref{eq:asympt_derivative_general}) with the help of (\ref{eq:Bessel_def2}). 
   
   Using  (\ref{eq:associated_legendre_def}), we have 
  \begin{equation} \label{eq:proof_lemma3_equivalence}
 P_{n+c}^m\left(\frac{1+x/n^2}{1-x/n^2}\right)=(-1)^m\left(-\frac{4x/n^2}{(1-x/n^2)^2}\right)^{m/2}\frac{d^m}{dy^m}P_{n+c}(y)\Big\vert_{y=\frac{1+x/n^2}{1-x/n^2}}
\end{equation}
where 
\begin{equation} \label{eq:first_term_asympt}
  \left(-\frac{4x/n^2}{(1-x/n^2)^2}\right)^{m/2}=\left(\frac{\iota}{n}\right)^{m}(4x)^{m/2}\left(1+\mathcal{O}\left(\frac{1}{n^2}\right)\right).
\end{equation}
Combining the limiting properties of (\ref{eq:asympt_derivative2}) and (\ref{eq:first_term_asympt}), we obtain the result (\ref{eq:Watson_modif_transf_general}).\qed
\end{proofof}

The second term of the expansion (\ref{eq:asympt_derivative_general}) depends on a modified Bessel function of one order less than that of the leading term. This is in contrast to the second term of the expansion (\ref{eq:asympt_edelman}), which is given by a modified Bessel function of two order less than that of the leading term.

As a by-product of Lemma \ref{Lemma_legendre_correction},  we also give the following Corollary, which presents results that have not been reported elsewhere, to the best of our knowledge\footnote{A similar result to (\ref{eq:Watson_modif_transf}) was presented in \cite[p. 156 eq. (3)]{WatsonBook} without proof. Although that result relates Bessel functions with associated Legendre polynomials when their arguments lie outside $[-1,1]$ as $n$ grows, it involves a different argument, omits the complex constant $\iota^m$ and is not valid for the whole range of values indicated in \cite{WatsonBook}. That result should instead read
   \begin{equation} 
 \lim_{n\rightarrow\infty} n^{m}P_n^{-m}\left(\cosh\left(\frac{x}{n}\right)\right)=\iota^m I_m(x).
\end{equation}.}.

\begin{corollary} \label{corollary_legendre}
For fixed $m\geq 0$ and $x>0$,
\begin{equation} \label{eq:asympt_derivative}
  \lim_{n\rightarrow\infty}n^{-2m}\frac{d^m}{dy^m}P_n(y)\bigg\vert_{y=\frac{1+x/n^2}{1-x/n^2}}=\frac{I_m(\sqrt{4x})}{(4x)^{m/2}}.
\end{equation}
For fixed $m\in\mathbb{Z}$ and $x\geq 0$,
\begin{equation}
  \begin{split}
  \label{eq:Watson_modif_transf}
  \lim_{n\rightarrow\infty}  n^{-m}P_n^{m}\left(\frac{1+x/n^2}{1-x/n^2}\right)=(-\iota)^m I_m(\sqrt{4x}).
  \end{split}
\end{equation} 
\end{corollary}
      
      \begin{remark} \label{remark2}
        Although Corollary \ref{corollary_legendre} presents asymptotic results for Legendre polynomials of degree $n$ when ${n\rightarrow\infty}$, they are also valid for Legendre polynomials of degree $n+c$ when $n\rightarrow\infty$, with fixed $c\in\mathbb{Z}$, as shown in Lemma \ref{Lemma_legendre_correction}. We will use this in the proof of Theorem \ref{Theorem_asymptoticcalpha1xalpha1}.
      \end{remark}

\section{Proof of Theorem \ref{Theorem_min_central}}
\label{Appendix:Theorem_min_central}

We make use of the following result:
\begin{lemma} 
  Let $w(y)$ be a non-negative function with all its moments finite and $t_l\in\mathbb{R}$ for all $l$, with $t_l\neq t_k$ for all $l\neq k$. Then \label{lemma_Mehta} \emph{\cite[eq.~(22.4.11)]{MehtaBook}},
  \begin{equation}
\label{eq:Mehta_result}
\underbrace{\int\ldots\int}_{n{\rm-fold}}\prod_{i<j}^{n}(y_i-y_j)^2\prod_{k=1}^{n}w(y_k)dy_k\prod_{l=1}^{m}(t_l-y_k)={K}\frac{\det[\pi_{n+i-1}(t_j)]_{i,j=1,\ldots,m}}{\prod_{i<j}^{m}(t_i-t_j)}
\end{equation}
where $K$ is a normalization constant and $\pi_n$ is the $n$th order monic polynomial orthogonal with respect to the weight function $w(y)$ in the integration interval.
\end{lemma}

First, we prove the result (\ref{eq:smallest_central}) for the smallest eigenvalue. For the most part, the proof follows the strategy of \cite[Appendix A]{Pratha2013Three}, which considered the smallest eigenvalue distribution of the non-central complex Wishart model with rank-$1$ mean. 

We start by writing
\begin{equation} \label{eq:smallest_1-P_phi_n}
F_{\phi_n}(\xi)=\mathbb{P}(\phi_n<\xi)=1-\mathbb{P}(\phi_n\geq \xi)
\end{equation}
where $0\leq \xi< 1$ and
\begin{equation}
\mathbb{P}(\phi_n\geq \xi)=C\int_{\xi\leq x_n\leq\ldots\leq x_1\leq 1}\prod_{i<j}^n(x_i-x_j)^2\prod_{\substack{k=1}}^{n}x_k^{\alpha_1}(1-x_k)^{\alpha_2}dx_k
\end{equation}
with $C$  the normalization constant of the eigenvalue JPDF in (\ref{eq:JPDF_Beta}).
Since the integrand is symmetric in $x_1,\ldots,x_n$, we may write
\begin{equation} \label{eq:multiple_integral}
\mathbb{P}(\phi_n\geq \xi)=\frac{C}{n!}\underbrace{\int_{\xi}^1\ldots\int_{\xi}^1}_{n\text{-fold}}\prod_{i<j}^n(x_i-x_j)^2\prod_{\substack{k=1}}^{n}x_k^{\alpha_1}(1-x_k)^{\alpha_2}dx_k.
\end{equation}
After the multiple change of variables $y_i=(x_i-\xi)/(1-\xi)$, for $1\leq i\leq n$, we have
\begin{equation} \label{eq:multiple_integral2}
  \begin{split}
\mathbb{P}(\phi_n\geq \xi)=&\frac{C(1-\xi)^{n\alpha_1+n\alpha_2+n^2}}{n!}
\underbrace{\int_{0}^1\ldots\int_{0}^1}_{n\text{-fold}}\prod_{i<j}^n(y_i-y_j)^2\prod_{\substack{k=1}}^{n}\left(y_k+\frac{\xi}{1-\xi}\right)^{\alpha_1}(1-y_k)^{\alpha_2}dy_k.
\end{split}
\end{equation}
Rearranging the expression yields
\begin{equation} \label{eq:P_phi_n}
\mathbb{P}(\phi_n\geq \xi)=\frac{(-1)^{n\alpha_1}C(1-\xi)^{n\alpha_1+n\alpha_2+n^2}}{n!}T_n^{\alpha_1,\alpha_2}\left(-\frac{\xi}{1-\xi},1\right)
\end{equation}
where
\begin{equation} \label{eq:integral_T}
T_n^{\alpha_1,\alpha_2}(\beta,\gamma)=\underbrace{\int_0^1\ldots\int_0^1}_{n\text{-fold}}\prod_{i<j}^{n}(y_i-y_j)^2\prod_{k=1}^{n}dy_k\prod_{l=1}^{\alpha_1+\alpha_2}(t_l-y_k)
\end{equation}
with
\begin{equation} \label{eq:tl}
t_l=\begin{cases}
\beta, &  l=1,\ldots,\alpha_1\\
\gamma, & l=\alpha_1+1,\ldots,\alpha_1+\alpha_2.
\end{cases}
\end{equation}
Note that this is of the same form as (\ref{eq:Mehta_result}) in Lemma \ref{lemma_Mehta}; however, we cannot apply the lemma directly since $t_l$, $l=1,\ldots,\alpha_1+\alpha_2$, are not distinct. To proceed, first recognize that if $t_l$ for all $l$ were distinct, then
\begin{equation} \label{eq:after_Mehta}
\underbrace{\int_0^1\ldots\int_0^1}_{n\text{-fold}}\prod_{i<j}^{n}(y_i-y_j)^2\prod_{k=1}^{n}dy_k\prod_{l=1}^{\alpha_1+\alpha_2}(t_l-y_k)
=\widetilde{C}\frac{\det[\widetilde{P}_{n+i-1}(t_j)]_{i,j=1,\ldots,\alpha_1+\alpha_2}}{\prod_{i<j}^{\alpha_1+\alpha_2}(t_i-t_j)}
\end{equation}
where $\widetilde{C}$ is a normalization constant and $\widetilde{P}_{\nu}(x)$ is the $\nu$th order polynomial orthogonal with respect to $1$ in $[0,1]$. This is precisely the shifted Legendre polynomial, defined in Section \ref{section:legendre}. Our desired integral in (\ref{eq:integral_T}) can be evaluated from (\ref{eq:after_Mehta}) by taking  limits as
\begin{align} \label{eq:T_n_det}
T_n^{\alpha_1,\alpha_2}(\beta,\gamma)= \widetilde{C}\lim_{\substack{t_1,\ldots,t_{\alpha_1}\rightarrow \beta\\ t_{\alpha_1+1},\ldots,t_{\alpha_1+\alpha_2}\rightarrow \gamma}}\frac{\det[\widetilde{P}_{n+i-1}(t_j)]_{\substack{i,j=1,\ldots,\alpha_1+\alpha_2}}}{\prod_{i<j}^{\alpha_1+\alpha_2}(t_i-t_j)}.
\end{align}
To evaluate these limits, we apply \cite[Lemma 2]{Chiani2010} and we have
\begin{equation}
\label{eq:T_n_simple}
T_n^{\alpha_1,\alpha_2}(\beta,\gamma)= \widetilde{C}^{\alpha_1,\alpha_2}(\beta,\gamma)\det\left(\mathbf{\Upsilon(\beta,\gamma)}\right)
\end{equation}
where $\mathbf{\Upsilon}(\beta,\gamma)$ is a $(\alpha_1+\alpha_2)\times(\alpha_1+\alpha_2)$ matrix defined by
\begin{equation}
  \mathbf{\Upsilon}(\beta,\gamma)=\left[\mathbf{D}_{\alpha_1}(\beta) \quad \mathbf{D}_{\alpha_2}(\gamma)\right]
\end{equation}
with $\mathbf{D}_{\alpha}(y)$ a $(\alpha_1+\alpha_2)\times\alpha$ matrix with entries
 \begin{equation} \label{eq:deriv_shift}
   \left[\mathbf{D}_{\alpha}(y)\right]_{ij}=\frac{d^{j-1}}{dy^{j-1}}\widetilde{P}_{n+i-1}(y)
 \end{equation}
 and
 \begin{equation}
\widetilde{C}^{\alpha_1,\alpha_2}(\beta,\gamma)=\frac{\widetilde{C}\left(\prod_{i=1}^{\alpha_1-1}i!\prod_{j=1}^{\alpha_2-1}j!\right)^{-1}}{(\gamma-\beta)^{\alpha_1\alpha_2}}.
\end{equation}
The product $\prod_{i=1}^{\alpha-1}i!$ is taken as $1$ when $\alpha=0$.

Using (\ref{eq:T_n_simple}), (\ref{eq:P_phi_n}) and (\ref{eq:smallest_1-P_phi_n}), we obtain
\begin{equation} \label{eq:appendix_resultproof1}
\begin{split}
  F_{\phi_n}(\xi)&=1-\widehat{C}^{-1}(1-\xi)^{\alpha_1\alpha_2+n\alpha_1+n\alpha_2+n^2}\det\left(\mathbf{\Upsilon}\left(-\frac{\xi}{1-\xi},1\right)\right).
\end{split}
\end{equation}
Since $F_{\phi_n}(0)=0$, 
\begin{equation} 
  \widehat{C}=\det\left(\mathbf{\Upsilon}\left(0,1\right)\right).
\end{equation}
With the help of (\ref{eq:def_shifted}) and the chain rule, we write the derivatives of the shifted Legendre polynomials in (\ref{eq:deriv_shift}) in terms of standard Legendre ones as
\begin{equation}
  \frac{d^k}{dz^k}\widetilde{P}_n(z)=2^k\frac{d^k}{dy^k}P_n(y)\bigg\vert_{y=2z-1},
\end{equation}
which gives the result for the smallest eigenvalue distribution in (\ref{eq:smallest_central}).  The result (\ref{eq:largest_central}) follows from (\ref{eq:smallest_central}) by applying the transformation in Remark \ref{remark1}.

\section{Proof of Proposition \ref{Theorem_min_central_Jacobi}}
\label{proof_prop1}
From \cite[eq. (3.16)]{Borodin2003} and \cite{Kaneko1993},
\begin{equation}
\label{eq:smallest_central_Forrester}
F_{\phi_n}(\xi)=h_{\alpha_1,\alpha_2}(\xi) 
\end{equation}
where\footnote{As mentioned in \cite{Borodin2003}, one can write $h_{\alpha_1,\alpha_2}(\xi)$ in terms of a polynomial in $\xi/(1-\xi)$. However, this polynomial is difficult to compute since it involves a sum over all partitions $\kappa$ of $k\in\{0,1,\ldots,n\alpha_1\}$ into no more than $\alpha_1$ parts.}
\begin{equation} \label{eq:h_Forrester}
h_{\alpha_1,\alpha_2}(\xi)=1-(1-\xi)^{n^2+n\alpha_1+n\alpha_2}\ _2\widetilde{F}_1(-n,n+\alpha_1+\alpha_2;\alpha_1;s_1,\ldots,s_{\alpha_1})\vert_{s_1=\ldots=s_{\alpha_1}=-\xi/(1-\xi)}
\end{equation}
with $_2\widetilde{F}_1$ the $\alpha_1$-dimensional complex Gauss hypergeometric function.
  First recognize that if $s_j$, $j=1,\ldots,\alpha_1$, were distinct in (\ref{eq:h_Forrester}), then \cite[eq.~(2.9)]{Richards}
  \begin{equation} \label{eq:h_ratio}
    _2\widetilde{F}_1(-n,n+\alpha_{1,2};\alpha_1;s_1,\ldots,s_{\alpha_1})=\frac{\det\left[s_j^{\alpha_1-i}\ _2\mathcal{F}_1(-n-i+1,n+\alpha_{1,2}-i+1:\alpha_1-i+1;s_j)\right]_{ij=1,\ldots,\alpha_1}}{\prod_{i<j}^{\alpha_1}(s_i-s_j)}
  \end{equation}
  where $\alpha_{1,2}=\alpha_1+\alpha_2$ and $_2\mathcal{F}_1$ is the Gauss hypergeometric function of scalar argument. Our desired expression in (\ref{eq:h_Forrester}) can be evaluated from (\ref{eq:h_ratio}) by taking limits as
  \begin{equation}
  \begin{split}
    h_{\alpha_1,\alpha_2}(\xi)=1-&(1-\xi)^{n^2+n\alpha_1+n\alpha_2}\\
    &\times\lim_{s_1,\ldots,s_{\alpha_1}\rightarrow -\xi/(1-\xi)}\frac{\det\left[s_j^{\alpha_1-i}\ _2\mathcal{F}_1(-n-i+1,n+\alpha_{1,2}-i+1:\alpha_1-i+1;s_j)\right]_{ij=1,\ldots,\alpha_1}}{\prod_{i<j}^{\alpha_1}(s_i-s_j)}
    \end{split}
  \end{equation}
  where the Gauss hypergeometric function of scalar argument can be expressed in terms of Jacobi polynomials as \cite[eq.~(15.4.6)]{Abramowitz}
  \begin{equation}
    _2\mathcal{F}_1(-n-i+1,n+\alpha_{1,2}-i+1:\alpha_1-i+1;s_j)=\frac{(n+i-1)!(\alpha_1-i)!}{(\alpha_1+n-1)!}P_{n+i-1}^{(\alpha_1-i,\alpha_2-i+1)}(1-2s_j).
  \end{equation}
  
  To evaluate these limits, we apply \cite[Lemma 2]{Chiani2010} and we have
  \begin{equation} \label{eq:proof_Jacobi}
  \begin{split}
    h_{\alpha_1,\alpha_2}(\xi)=1-&\prod_{k=1}^{\alpha_1}\frac{(n+k-1)!(\alpha_1-k)!}{(\alpha_1+n-1)!(k-1)!}(1-\xi)^{n^2+n\alpha_1+n\alpha_2}\\
    &\times\det\left[\frac{d^{\alpha_1-j}}{dy^{\alpha_1-j}}\left[y^{\alpha_1-i} P_{n+i-1}^{(\alpha_1-i,\alpha_2-i+1)}(1-2y)\right]\Big\vert_{y=-\xi/(1-\xi)}\right]_{i,j=1,\ldots,\alpha_1}.
        \end{split}
  \end{equation}
  Finally, we obtain the result by applying the Leibniz rule to the entries of the determinant. 

\section{Proof of Theorem \ref{Theorem_asymptoticcalpha1xalpha1}}
\label{Appendix:Asymptoticalpha1xalpha1}

As before, we prove the result (\ref{eq:asympt_min}) for the smallest eigenvalue, with the result (\ref{eq:asympt_max}) then following from  Remark \ref{remark1}.

First consider the case $\alpha_2=0$. Applying (\ref{eq:legendre_minus_argument}) to the entries of $\mathbf{E}_{\alpha_1}(y)$ and some algebraic simplifications, we obtain
\begin{equation} \label{eq:appendix_alpha2_0}
  \begin{split}
  &F_{n^2\phi_n}(x)=F_{\phi_n}(x/n^2)=1-(1-x/n^2)^{n^2+\alpha_1 n}\frac{\det\left(\mathbf{E}_{\alpha_1}(x^\star)\right)}{\det(\mathbf{E}_{\alpha_1}(1))}
\end{split}
\end{equation}
where $x^\star=(1+x/n^2)/(1-x/n^2)$.

If one takes $n$ large and applies Corollary \ref{corollary_legendre} to the entries of the numerator determinant of (\ref{eq:appendix_alpha2_0}), given in (\ref{eq:E}), we obtain
\begin{equation}
  \left[\mathbf{E}_{\alpha_1}(x^\star)\right]_{ij}=n^{2(j-1)}\frac{I_{j-1}(\sqrt{4x})}{(4x)^{(j-1)/2}}+o\left(n^{2(j-1)}\right),
\end{equation}
while for the denominator, with (\ref{eq:g}), we obtain
\begin{equation} \label{eq:appendix_E(1)_asympt}
  \left[\mathbf{E}_{\alpha_1}(1)\right]_{ij}=n^{2(j-1)}\frac{2^{1-j}}{(j-1)!}+o\left(n^{2(j-1)}\right).
\end{equation}
Hence, replacing the entries of both determinants with their leading order terms gives clearly a $0/0$ indetermination in (\ref{eq:appendix_alpha2_0}). To circumvent this, we iteratively make a set of manipulations to the determinant of $\mathbf{E}_{\alpha_1}(y)$ by using properties of Legendre polynomials (Lemmas \ref{property2} and \ref{Lemma_legendre2} and Corollary \ref{corollary_property2}), following a similar method as in \cite{Forrester1994alpha_det} for the Laguerre case. In particular, we iteratively make row operations and use the recurrence property in Corollary \ref{corollary_property2} to modify the derivative orders of the entries within a specific column, so that when $y=x^\star$, by virtue of Corollary \ref{corollary_legendre}, they approach a different Bessel function in the limit, which avoids the $0/0$ indetermination. However, the recurrence property in Corollary \ref{corollary_property2} for the Legendre case presents a certain range of validity, which will prevent its application for some entries. Also, contrary to the recurrence property of Laguerre polynomials used in \cite{Forrester1994alpha_det}, having the constant $(n-m+1)$ on the left-hand side of Corollary \ref{corollary_property2} makes the derivation more cumbersome. We first demonstrate the result for the case $\alpha_1=3$, in order to shed light on the set of manipulations required to prove the general result.

\subsection{An illustrative case: $\alpha_1=3$ and $\alpha_2=0$} \label{proof_3x3}

This corresponds to the simplest case that demonstrates the challenge posed by the recurrence property of Corollary \ref{corollary_property2} or, more generally, Lemma \ref{property2}.
For $\alpha_1=3$ and $\alpha_2=0$, we have
\begin{align}
\mathbf{E}_3(y)= &\begin{bmatrix}
    P_n\left(y\right) & \frac{dP_n\left(y\right)}{dy}  &  \frac{d^{2}P_n\left(y\right)}{dy^{2}}\\
    P_{n+1}\left(y\right) & \frac{dP_{n+1}\left(y\right)}{dy}  &  \frac{d^{2}P_{n+1}\left(y\right)}{dy^{2}}\\
    P_{n+2}\left(y\right) & \frac{dP_{n+2}\left(y\right)}{dy}  &  \frac{d^{2}P_{n+2}\left(y\right)}{dy^{2}}\\
      \end{bmatrix}.
            \end{align}            
When $y=x^\star$, by virtue of Corollary \ref{corollary_legendre}, we identify for large $n$
\begin{align} \label{eq:E3_orders}
\mathbf{E}_3(x^\star)= &\begin{bmatrix}
    \vert & \vert  &  \vert\\
    \mathcal{O}(1) & \mathcal{O}(n^2)  &  \mathcal{O}(n^4)\\
    \vert & \vert  &  \vert\\
      \end{bmatrix}.
            \end{align}
We apply a set of iterative operations which will successively decrease the order (in $n$) from one row to the next in (\ref{eq:E3_orders}). This will make use of the recurrence properties of Legendre polynomials in Lemma \ref{property2} and Corollary \ref{corollary_property2}. Although we could apply the more general Lemma \ref{property2} instead, Corollary \ref{corollary_property2} will be useful to illustrate the purpose of each iteration. In the first iteration, to facilitate the application of Corollary \ref{corollary_property2}, we scale the third row of $\mathbf{E}_{3}(y)$ by $y$ and then subtract the second row. We then scale the second row by $y$ and then subtract the first row. Note that this does not alter the first row. This procedure yields $\det(\mathbf{E}_3(y))=y^{-2}\det(\mathbf{\Theta}_n^{(1)}(y))$ with
\begin{equation}
\mathbf{\Theta}^{(1)}_n(y)= \begin{bmatrix}
    P_n\left(y\right) & \frac{dP_n\left(y\right)}{dy}  &  \frac{d^{2}P_n\left(y\right)}{dy^{2}}\\
    yP_{n+1}\left(y\right)- P_n\left(y\right) & y\frac{dP_{n+1}\left(y\right)}{dy}-\frac{dP_n\left(y\right)}{dy}  &  y\frac{d^{2}P_{n+1}\left(y\right)}{dy^{2}}-\frac{d^{2}P_n\left(y\right)}{dy^{2}}\\
    yP_{n+2}\left(y\right)-P_{n+1}\left(y\right) & y\frac{dP_{n+2}\left(y\right)}{dy}-\frac{dP_{n+1}\left(y\right)}{dy}  &  y\frac{d^{2}P_{n+2}\left(y\right)}{dy^{2}}-\frac{d^{2}P_{n+1}\left(y\right)}{dy^{2}}\\
      \end{bmatrix}.
\end{equation}

We can now apply Corollary \ref{corollary_property2} to the modified entries of the second and third columns. For the modified entries of the first column, we use their Rodrigues' formula representation (\ref{eq:derivative_Legendre_binomial}) and then employ Lemma \ref{property2}. This leads to
\begin{equation}
\mathbf{\Theta}^{(1)}_n(y)=\begin{bmatrix}
    P_n\left(y\right) & \frac{dP_n\left(y\right)}{dy}   &  \frac{d^{2}P_n\left(y\right)}{dy^{2}}\\
   \frac{(n+2)}{2^{n+1}(n+1)!}\frac{d^{n}}{dy^n}\left[(y^2-1)^{n+1}\right]  & (n+1)P_{n+1}(y)  & n\frac{dP_{n+1}(y)}{dy}\\
   \frac{(n+3)}{2^{n+2}(n+2)!}\frac{d^{n+1}}{dy^{n+1}}\left[(y^2-1)^{n+2}\right]  & (n+2)P_{n+2}(y)  & (n+1)\frac{dP_{n+2}(y)}{dy}\\
    \end{bmatrix},
\end{equation}
concluding the first iteration. In the second iteration, we repeat the same manipulations, but this time we only scale the third row by $y$ and subtract the second row. This gives $\det(\mathbf{E}_3(y))=y^{-3}\det(\mathbf{\Theta}_n^{(2)}(y))$ with
\begin{equation}
\mathbf{\Theta}^{(2)}_n(y)=\begin{bmatrix}
    P_n\left(y\right) & \frac{dP_n\left(y\right)}{dy}   &  \frac{d^{2}P_n\left(y\right)}{dy^{2}}\\
   \frac{(n+2)}{2^{n+1}(n+1)!}\frac{d^{n}}{dy^n}\left[(y^2-1)^{n+1}\right]  & (n+1)P_{n+1}(y)  & n\frac{dP_{n+1}(y)}{dy}\\
   a_{1,1}(y)  & a_{1,2}(y)  & a_{1,3}(y)\\
    \end{bmatrix}
\end{equation}
where
\begin{align}
  {a}_{1,1}(y)=\ &(n+3)\frac{y}{2^{n+2}(n+2)!}\frac{d^{n+1}}{dy^{n+1}}\left[(y^2-1)^{n+2}\right]-(n+2)\frac{1}{2^{n+1}(n+1)!}\frac{d^{n}}{dy^{n}}\left[(y^2-1)^{n+1}\right]\\
  a_{1,2}(y)=\ &(n+2)y P_{n+2}(y)-(n+1)P_{n+1}(y),\\
  a_{1,3}(y)=\ &(n+1)y\frac{d}{dy}P_{n+2}(y)+n\frac{d}{dy}P_{n+1}(y).
\end{align}
We then employ Lemma \ref{property2} and Corollary \ref{corollary_property2} to rewrite the entries $a_{1,j}(y)$. Specifically, we rewrite $a_{1,1}(y)$ as
\begin{equation}
  \begin{split}
  {a}_{1,1}(y)=\ &(n+3)\left[\frac{y}{2^{n+2}(n+2)!}\frac{d^{n+1}}{dy^{n+1}}\left[(y^2-1)^{n+2}\right]-\frac{n+2}{n+3}\frac{1}{2^{n+1}(n+1)!}\frac{d^{n}}{dy^{n}}\left[(y^2-1)^{n+1}\right]\right]\\
  =\ &(n+3)\left[\frac{y}{2^{n+2}(n+2)!}\frac{d^{n+1}}{dy^{n+1}}\left[(y^2-1)^{n+2}\right]-\left(1-\frac{1}{n+3}\right)\frac{1}{2^{n+1}(n+1)!}\frac{d^{n}}{dy^{n}}\left[(y^2-1)^{n+1}\right]\right]\\
  =\ &\frac{(n+3)(n+4)}{2^{n+2}(n+2)!}\frac{d^{n}}{dy^{n}}\left[(y^2-1)^{n+2}\right]+\frac{1}{2^{n+1}(n+1)!}\frac{d^{n}}{dy^{n}}\left[(y^2-1)^{n+1}\right]
  \end{split}
\end{equation}
where the first term of the last line followed from Lemma \ref{property2}. The entries $a_{1,2}(y)$ and $a_{1,3}(y)$ are handled similarly, by employing Lemma \ref{property2} and Corollary \ref{corollary_property2} respectively, giving
\begin{align}
  a_{1,2}(y)=\ &\frac{(n+2)(n+3)}{2^{n+2}(n+2)!}\frac{d^{n+1}}{dy^{n+1}}\left[(y^2-1)^{n+2}\right]+P_{n+1}(y),\\
  a_{1,3}(y)=\ &(n+1)(n+2)P_{n+2}(y)+\frac{d}{dy}P_{n+1}(y).
\end{align}
At this point, we apply Lemma \ref{Lemma_legendre2} to the entries below the main diagonal of $\mathbf{\Theta}^{(2)}_n(y)$ to obtain
\begin{equation}
  \begin{split}
&\mathbf{\Theta}^{(2)}_n(y)=\\
&\begin{bmatrix}
    P_n\left(y\right) & \frac{dP_n\left(y\right)}{dy}   &  \frac{d^{2}P_n\left(y\right)}{dy^{2}}\\
   \frac{\sqrt{1-y^2}}{n+1}P_{n+1}^1(y)  & (n+1)P_{n+1}(y)  & n\frac{dP_{n+1}(y)}{dy}\\
   \frac{1-y^2}{(n+1)(n+2)}P_{n+2}^2(y)+\frac{\sqrt{1-y^2}}{(n+1)(n+2)}P_{n+1}^1(y)  & \sqrt{1-y^2}P_{n+2}^1(y)+P_{n+1}(y) &(n+1)(n+2)P_{n+2}(y)+\frac{dP_{n+1}(y)}{dy}
    \end{bmatrix}.
    \end{split}
\end{equation}
Note that the entries of the third row of $\mathbf{\Theta}^{(2)}_n(y)$ have an additional term with respect to the previous rows as a result of the manipulations. This is due to the fact that the recurrence formula of Legendre polynomials in Lemma \ref{property2} (or Corollary \ref{corollary_property2}) has the constant factor $(n-m+1)$ in its left-hand side. We will need to consider such additional terms when taking limits.

When $y=x^\star$, $1-y^2=-4x/n^2$. With this in mind, by virtue of Corollary \ref{corollary_legendre}, we now have
\begin{equation} \label{eq:E3_orders_after}
\mathbf{\Theta}^{(2)}_n(x^\star)=
\begin{bmatrix}
    \mathcal{O}(1) & \mathcal{O}(n^2)   &  \mathcal{O}(n^4)\\
   \mathcal{O}(1/n)  & \mathcal{O}(n)  & \mathcal{O}(n^3)\\
   \mathcal{O}(1/n^2)  & \mathcal{O}(1) & \mathcal{O}(n^2)
    \end{bmatrix},
\end{equation}
where in contrast to (\ref{eq:E3_orders}), as alluded earlier, the order in $n$ has been successively reduced in the second and third rows. This is a consequence of iteratively applying Corollary \ref{corollary_property2} or, more generally, \mbox{Lemma \ref{property2}}. Effectively, when applying Corollary \ref{corollary_property2} to the modified upper-triangular entries, the order in $n$ is reduced by one. This can be seen from the reduction in the derivative order of the Legendre polynomials, which reduces the order in $n$ by two (Corollary \ref{corollary_legendre}), and from the factor $(n-m+1)$ (left-hand side of Corollary \ref{corollary_property2}), which increases that order by one. The same effect occurs when applying Lemma \ref{property2} to the lower-triangular entries, even though Lemma \ref{property2} does not explicitly show this order reduction.

Next, we perform some manipulations in order to apply the limiting results of Corollary \ref{corollary_legendre}, while making the entries all of order $1$. We divide the $j$th column by $n^{j-1}(1-y^2)^{(3-j)/2}$ for $j=1,\ldots,3$, and multiply the $i$th row by $(1-y^2)^{(3-i)/2}$ for $i=1,\ldots,3$. This gives $\det(\mathbf{E}_3(y)) = (y/n)^{-3}\det(\mathbf{\tilde\Theta}_n^{(2)}(y))$ with
\begin{equation}
  \begin{split}
\mathbf{\tilde\Theta}^{(2)}_n(&y)=\\
&\begin{bmatrix}
    P_n\left(y\right) & \frac{\sqrt{1-y^2}}{n}\frac{dP_n\left(y\right)}{dy}   &  \frac{1-y^2}{n^2}\frac{d^{2}P_n\left(y\right)}{dy^{2}}\\
   \frac{P_{n+1}^1(y)}{n+1}  & \frac{n+1}{n}P_{n+1}(y)  & \frac{\sqrt{1-y^2}}{n}\frac{dP_{n+1}(y)}{dy}\\
   \frac{P_{n+2}^2(y)}{(n+1)(n+2)}+\frac{(1-y^2)^{-1/2}}{(n+1)(n+2)}P_{n+1}^1(y)  & \frac{P_{n+2}^1(y)}{n}+\frac{P_{n+1}(y)}{n\sqrt{1-y^2}}  & \frac{(n+1)(n+2)}{n^2}P_{n+2}(y)+\frac{1}{n^2}\frac{dP_{n+1}(y)}{dy}
    \end{bmatrix}.
    \end{split}
\end{equation}

At this point, by virtue of Corollary \ref{corollary_legendre}, the columns of $\mathbf{\tilde\Theta}^{(2)}_n(x^\star)$ are linearly independent when $n\rightarrow\infty$. We then rewrite (\ref{eq:appendix_alpha2_0}) for $\alpha_1=3$ and $\alpha_2=0$ as
\begin{equation} \label{eq:equiv_det2_alpha3}
  \begin{split}
  &F_{\phi_n}(x/n^2)=1-\frac{(1-x/n^2)^{n^2+3n+3}}{(1+x/n^2)^{3}}\frac{\det\left({\mathbf{\tilde\Theta}}^{(2)}_n(x^\star)\right)}{\det\left({\mathbf{\tilde\Theta}}^{(2)}_n(1)\right)}
  \end{split}
\end{equation}
and take the $n\rightarrow\infty$ limit. Specifically, applying Corollary \ref{corollary_legendre} to the entries of $\mathbf{\tilde\Theta}^{(2)}_n(x^\star)$, and recalling that $1-y^2=-4x/n^2$ when $y=x^\star$, we obtain that $\lim_{n\rightarrow\infty}\det({\mathbf{\tilde\Theta}}^{(2)}_n(x^\star))=\det\left(\mathbf{L}^{(0)}(x)\right)$ with
\begin{equation}
  \begin{split}
  \mathbf{L}^{(0)}(x)=\begin{bmatrix}
   I_0(\sqrt{4x}) & \iota I_1(\sqrt{4x}) & -I_2(
   \sqrt{4x}) \\
    -\iota I_1(\sqrt{4x}) & I_0(\sqrt{4x}) & \iota I_1(\sqrt{4x}) \\
    -I_2(\sqrt{4x})-\frac{I_1(\sqrt{4x})}{\sqrt{4x}} & -\iota I_1(\sqrt{4x})-\iota\frac{I_0(\sqrt{4x})}{\sqrt{4x}} & I_0(\sqrt{4x})+\frac{I_1(\sqrt{4x})}{\sqrt{4x}}
\end{bmatrix}.
\end{split}
\end{equation}
To make all complex constants vanish, we divide the $j$th column by $(-\iota)^{3-j}$ for $j=1,\ldots,3$ and then multiply the $i$th row by $(-\iota)^{3-i}$ for $i=1,\ldots,3$, so that $\det(\mathbf{L}^{(0)}(x))=\det(\mathbf{\tilde L}^{(0)}(x))$ with
\begin{equation}
  \begin{split}
  \mathbf{\tilde L}^{(0)}(x)=\begin{bmatrix}
   I_0(\sqrt{4x}) &  I_1(\sqrt{4x}) & I_2(
   \sqrt{4x}) \\
     I_1(\sqrt{4x}) & I_0(\sqrt{4x}) &  I_1(\sqrt{4x}) \\
    I_2(\sqrt{4x})+\frac{I_1(\sqrt{4x})}{\sqrt{4x}} &  I_1(\sqrt{4x})+\frac{I_0(\sqrt{4x})}{\sqrt{4x}} & I_0(\sqrt{4x})+\frac{I_1(\sqrt{4x})}{\sqrt{4x}}
\end{bmatrix}.
\end{split}
\end{equation}
Notice that we can simplify the determinant by subtracting the second row scaled by $(4x)^{-1/2}$ from the third row, so that $\det(\mathbf{L}^{(0)}(x))=\det(\mathbf{\tilde L}^{(1)}(x))$ with
\begin{equation} 
  \begin{split}
  \mathbf{\tilde L}^{(1)}(x)=\begin{bmatrix}
   I_0(\sqrt{4x}) &  I_1(\sqrt{4x}) & I_2(
   \sqrt{4x}) \\
     I_1(\sqrt{4x}) & I_0(\sqrt{4x}) &  I_1(\sqrt{4x}) \\
    I_2(\sqrt{4x}) &  I_1(\sqrt{4x}) & I_0(\sqrt{4x})
\end{bmatrix}.
\end{split}
\end{equation}

We then evaluate the $n\rightarrow\infty$ limit of (\ref{eq:equiv_det2_alpha3}) as
\begin{equation} \label{eq:equiv_det_alpha3_2}
  \lim_{n\rightarrow\infty}F_{\phi_n}(x/n^2)=1-e^{-x}\frac{\det\left(\mathbf{\tilde L}^{(1)}(x)\right)}{\det\left(\mathbf{\tilde L}^{(1)}(0)\right)}
\end{equation}
where we have used the limit definition of the exponential. Since $I_m(0)=0$ for all $m\in\mathbb{Z}^+$ and $I_0(0)=1$, explicit computation of the denominator determinant in (\ref{eq:equiv_det_alpha3_2}) gives $1$. Finally, noting that
\begin{equation}
\left[\mathbf{\tilde L}^{(1)}(x)\right]_{ij}=  I_{i-j}(\sqrt{4x})
\end{equation}
since $I_m(z)=I_{-m}(z)$ for all $m\in\mathbb{Z}$, we obtain the result (\ref{eq:asympt_min}) for $\alpha_1=3$ and $\alpha_2=0$.

\subsection{Proof for arbitrary $\alpha_1$ and $\alpha_2=0$}

For arbitrary $\alpha_1$ and $\alpha_2=0$, we use the same approach as in the case $\alpha_1=3$ and $\alpha_2=0$. First, we make row operations to $\mathbf{E}_{\alpha_1}(y)$ to successively reduce the    order in $n$ of entries of the same column, similar to (\ref{eq:E3_orders_after}). Then, we perform some manipulations to facilitate the application of Corollary \ref{corollary_legendre}. Finally, after taking $n\rightarrow\infty$, we perform some row operations to simplify the entries of the limiting determinant. 

In each iteration, for specified $k$ values, we successively scale the $k$th last row of $\mathbf{E}_{\alpha_1}(y)$ by $y$ and then subtract the $(k-1)$th last row to facilitate the application of the Legendre recurrence properties to the entries of the $k$th last row. In the first iteration, we perform those row operations for $k=1,2\ldots,\alpha_1-1$.  This does not alter the first row. Let $\tau=n+\alpha_1-1$. This procedure yields $\det(\mathbf{E}_{\alpha_1}(y))=y^{-\alpha_1+1}\det(\mathbf{\Theta}^{(1)}_n(y))$ with
\begin{equation}
  \label{eq:General_alpha1_det_1iteration}
  \begin{split}
\mathbf{\Theta}^{(1)}_n(y)=\begin{bmatrix}
    P_n\left(y\right) & \frac{dP_n\left(y\right)}{dy} & \ldots  &  \frac{d^{\alpha_1-1}P_n\left(y\right)}{dy^{\alpha_1-1}}\\
   \frac{n+2}{2^{n+1}(n+1)!}\frac{d^{n}}{dy^n}\left[(y^2-1)^{n+1}\right]  & (n+1)P_{n+1}(y) & \ldots & (n-\alpha_1+3)\frac{d^{\alpha_1-2}P_{n+1}\left(y\right)}{dy^{\alpha_1-2}}\\
   \frac{n+3}{2^{n+2}(n+2)!}\frac{d^{n+1}}{dy^{n+1}}\left[(y^2-1)^{n+2}\right]  & (n+2)P_{n+2}(y) & \ldots & (n-\alpha_1+4)\frac{d^{\alpha_1-2}P_{n+2}\left(y\right)}{dy^{\alpha_1-2}}\\
    \vdots & \vdots &  & \vdots\\
    \frac{\tau+1}{2^{\tau}\tau!}\frac{d^{\tau-1}}{dy^{\tau-1}}\left[(y^2-1)^{\tau}\right] & \tau P_{\tau}(y) & \ldots  &  (n+1)\frac{d^{\alpha_1-2}P_{\tau}\left(y\right)}{dy^{\alpha_1-2}}
    \end{bmatrix}
\end{split}
\end{equation}
where the modified entries of the first column followed  from the Rodrigues' formula (\ref{eq:derivative_Legendre_binomial}) and Lemma \ref{property2}, and the rest of modified entries followed from  Corollary \ref{corollary_property2}.

For the sake of notational simplicity, we unify the remaining iterative operations by only employing Lemma \ref{property2} to simplify the modified entries. As noted in the previous subsection, Corollary \ref{corollary_property2} allowed to better illustrate the purpose of each iteration. Here, using Corollary \ref{corollary_property2} produces cumbersome notation, and we will resort to the more general Lemma \ref{property2}. Then, we first apply the Rodrigues' formula (\ref{eq:derivative_Legendre_binomial}) to the modified entries beyond the first column of $\mathbf{\Theta}_n^{(1)}(y)$ to obtain
\begin{equation}
  \label{eq:General_alpha1_det_1iteration2}
  \begin{split}
\mathbf{\Theta}^{(1)}_n(&y)=\\ &\begin{bmatrix}
    P_n\left(y\right) & \frac{dP_n\left(y\right)}{dy} & \ldots  &  \frac{d^{\alpha_1-1}P_n\left(y\right)}{dy^{\alpha_1-1}}\\
   \frac{n+2}{2^{n+1}(n+1)!}\frac{d^{n}}{dy^n}\left[(y^2-1)^{n+1}\right]  & \frac{n+1}{2^{n+1}(n+1)!}\frac{d^{n+1}}{dy^{n+1}}\left[(y^2-1)^{n+1}\right] & \ldots & \frac{n-\alpha_1+3}{2^{n+1}(n+1)!}\frac{d^{\tau}}{dy^{\tau}}\left[(y^2-1)^{n+1}\right]\\
   \frac{n+3}{2^{n+2}(n+2)!}\frac{d^{n+1}}{dy^{n+1}}\left[(y^2-1)^{n+2}\right]  & \frac{n+2}{2^{n+2}(n+2)!}\frac{d^{n+2}}{dy^{n+2}}\left[(y^2-1)^{n+2}\right] & \ldots & \frac{n-\alpha_1+4}{2^{n+2}(n+2)!}\frac{d^{\tau+1}}{dy^{\tau+1}}\left[(y^2-1)^{n+2}\right]\\
    \vdots & \vdots &  & \vdots\\
    \frac{\tau+1}{2^{\tau}\tau!}\frac{d^{\tau-1}}{dy^{\tau-1}}\left[(y^2-1)^{\tau}\right] & \frac{\tau}{2^{\tau}\tau!}\frac{d^{\tau}}{dy^{\tau}}\left[(y^2-1)^{\tau}\right] & \ldots  &  \frac{n+1}{2^{\tau}\tau!}\frac{d^{\tau+\alpha_1-2}}{dy^{\tau+\alpha_1-2}}\left[(y^2-1)^{\tau}\right]
    \end{bmatrix},
\end{split}
\end{equation}
and we repeat the same row operations, but this time for ${k=1,\ldots,\alpha_1-2}$, so that $\det(\mathbf{E}_{\alpha_1}(y))=y^{-2\alpha_1+3}\det(\mathbf{\Theta}^{(2)}_n(y))$ \sloppy with 
\begin{equation} \label{eq:limit_before_eq}
  \begin{split}
  \mathbf{\Theta}^{(2)}_n(y)=\begin{bmatrix}
    P_n(y) & \frac{dP_n\left(y\right)}{dy} & \ldots  &  \frac{d^{\alpha_1-1}P_n\left(y\right)}{dy^{\alpha_1-1}}\\
    \frac{n+2}{2^{n+1}(n+1)!}\frac{d^{n}}{dy^n}\left[(y^2-1)^{n+1}\right]  & \frac{n+1}{2^{n+1}(n+1)!}\frac{d^{n+1}}{dy^{n+1}}\left[(y^2-1)^{n+1}\right] & \ldots & \frac{n-\alpha_1+3}{2^{n+1}(n+1)!}\frac{d^{\tau}}{dy^{\tau}}\left[(y^2-1)^{n+1}\right]\\
    {a}^{(2)}_{1,1}(y) & {a}^{(2)}_{1,2}(y)  & \ldots & {a}^{(2)}_{1,\alpha_1}(y)\\
    {a}^{(2)}_{2,1}(y) & {a}^{(2)}_{2,2}(y)  & \ldots & {a}^{(2)}_{2,\alpha_1}(y)\\
    \vdots & \vdots &  & \vdots\\
    {a}^{(2)}_{\alpha_1-2,1}(y) & {a}^{(2)}_{\alpha_1-2,2}(y)  & \ldots & {a}^{(2)}_{\alpha_1-2,\alpha_1}(y)\\
\end{bmatrix}
\end{split}
\end{equation}
where
\begin{equation} \label{eq:aij_first_iter}
  a_{i,j}^{(2)}(y)=\frac{(n+i-j+3)(n+i-j+4)}{2^{n+i+1}(n+i+1)!}\frac{d^{n+i+j-2}}{dy^{n+i+j-2}}\left[(y^2-1)^{n+i+1}\right]+\frac{1}{2^{n+i}(n+i)!}\frac{d^{n+i+j-2}}{dy^{n+i+j-2}}\left[(y^2-1)^{n+i}\right]
\end{equation}
with the first term of (\ref{eq:aij_first_iter}) following from Lemma \ref{property2}.
Like in the case $\alpha_1=3$ and $\alpha_2=0$, the second iteration gives an additional term for the entries below the second row. In the third iteration, we repeat the same procedure, but this time for $k=1,\ldots,\alpha_1-3$, obtaining $\det(\mathbf{E}_{\alpha_1}(y))=y^{-3\alpha_1+6}\det(\mathbf{\Theta}^{(3)}_n(y))$ with
\begin{equation} 
  \begin{split}
  \mathbf{\Theta}^{(3)}_n(y)=\begin{bmatrix}
    P_n(y) & \frac{dP_n\left(y\right)}{dy} & \ldots  &  \frac{d^{\alpha_1-1}P_n\left(y\right)}{dy^{\alpha_1-1}}\\
    \frac{n+2}{2^{n+1}(n+1)!}\frac{d^{n}}{dy^n}\left[(y^2-1)^{n+1}\right]  & \frac{n+1}{2^{n+1}(n+1)!}\frac{d^{n+1}}{dy^{n+1}}\left[(y^2-1)^{n+1}\right] & \ldots & \frac{n-\alpha_1+3}{2^{n+1}(n+1)!}\frac{d^{\tau}}{dy^{\tau}}\left[(y^2-1)^{n+1}\right]\\
    {a}^{(2)}_{1,1}(y) & {a}^{(2)}_{1,2}(y)  & \ldots & {a}^{(2)}_{1,\alpha_1}(y)\\
    {a}^{(3)}_{2,1}(y) & {a}^{(3)}_{2,2}(y)  & \ldots & {a}^{(3)}_{2,\alpha_1}(y)\\
    {a}^{(3)}_{3,1}(y) & {a}^{(3)}_{3,2}(y)  & \ldots & {a}^{(3)}_{3,\alpha_1}(y)\\
    \vdots & \vdots &  & \vdots\\
    {a}^{(3)}_{\alpha_1-2,1}(y) & {a}^{(3)}_{\alpha_1-2,2}(y)  & \ldots & {a}^{(3)}_{\alpha_1-2,\alpha_1}(y)\\
\end{bmatrix}
\end{split}
\end{equation}
where
\begin{align}
  {a}^{(3)}_{i,j}(y)=\ &\frac{(n+i-j+3)(n+i-j+4)(n+i-j+5)}{2^{n+i+1}(n+i+1)!}\frac{d^{n+i+j-3}}{dy^{n+i+j-3}}\left[(y^2-1)^{n+i+1}\right]\nonumber\\
  &+\frac{3(n+i-j+3)}{2^{n+i}(n+i)!}\frac{d^{n+i+j-3}}{dy^{n+i+j-3}}\left[(y^2-1)^{n+i}\right].
\end{align}
In the fourth iteration, we repeat the same steps, but this time for $k=1,\ldots,\alpha_1-4$, to obtain $\det(\mathbf{E}_{\alpha_1}(y))=y^{-4\alpha_1+10}\det(\mathbf{\Theta}^{(4)}_n(y))$ with
\begin{equation} 
  \begin{split}
  \mathbf{\Theta}^{(4)}_n(y)=\begin{bmatrix}
    P_n(y) & \frac{dP_n\left(y\right)}{dy} & \ldots  &  \frac{d^{\alpha_1-1}P_n\left(y\right)}{dy^{\alpha_1-1}}\\
    \frac{n+2}{2^{n+1}(n+1)!}\frac{d^{n}}{dy^n}\left[(y^2-1)^{n+1}\right]  & \frac{n+1}{2^{n+1}(n+1)!}\frac{d^{n+1}}{dy^{n+1}}\left[(y^2-1)^{n+1}\right] & \ldots & \frac{n-\alpha_1+3}{2^{n+1}(n+1)!}\frac{d^{\tau}}{dy^{\tau}}\left[(y^2-1)^{n+1}\right]\\
    {a}^{(2)}_{1,1}(y) & {a}^{(2)}_{1,2}(y)  & \ldots & {a}^{(2)}_{1,\alpha_1}(y)\\
    {a}^{(3)}_{2,1}(y) & {a}^{(3)}_{2,2}(y)  & \ldots & {a}^{(3)}_{2,\alpha_1}(y)\\
    {a}^{(4)}_{3,1}(y) & {a}^{(4)}_{3,2}(y)  & \ldots & {a}^{(4)}_{3,\alpha_1}(y)\\
    {a}^{(4)}_{4,1}(y) & {a}^{(4)}_{4,2}(y)  & \ldots & {a}^{(4)}_{4,\alpha_1}(y)\\
    \vdots & \vdots &  & \vdots\\
     {a}^{(4)}_{\alpha_1-2,1}(y) & {a}^{(4)}_{\alpha_1-2,2}(y)  & \ldots & {a}^{(4)}_{\alpha_1-2,\alpha_1}(y)
\end{bmatrix}
\end{split}
\end{equation}
where
\begin{align} \label{eq:77}
  a_{i,j}^{(4)}(y)=\ &\frac{(n+i-j+3)(n+i-j+4)(n+i-j+5)(n+i-j+6)}{2^{n+i+1}(n+i+1)!}\frac{d^{n+i+j-4}}{dy^{n+i+j-4}}\left[(y^2-1)^{n+i+1}\right]\nonumber\\
  &+6\frac{(n+i-j+3)(n+i-j+4)}{2^{n+i}(n+i)!}\frac{d^{n+i+j-4}}{dy^{n+i+j-4}}\left[(y^2-1)^{n+i}\right]\nonumber\\
  &+\frac{3}{2^{n+i-1}(n+i-1)!}\frac{d^{n+i+j-4}}{dy^{n+i+j-4}}\left[(y^2-1)^{n+i-1}\right].
\end{align}
Observe that an additional term appears every two rows. After a total of $\alpha_1-1$ iterations, we obtain $\det(\mathbf{E}_{\alpha_1}(y))=y^{-\alpha_1(\alpha_1-1)/2}\det(\mathbf{\Theta}^{(\alpha_1-1)}_n(y))$ with
\begin{align}
  &\mathbf{\Theta}^{(\alpha_1-1)}_n(y)=\nonumber\\&\begin{bmatrix}
    P_n(y) & \frac{dP_n\left(y\right)}{dy} & \ldots  &  \frac{d^{\alpha_1-1}P_n\left(y\right)}{dy^{\alpha_1-1}}\\
    \frac{n+2}{2^{n+1}(n+1)!}\frac{d^{n}}{dy^n}\left[(y^2-1)^{n+1}\right]  & \frac{n+1}{2^{n+1}(n+1)!}\frac{d^{n+1}}{dy^{n+1}}\left[(y^2-1)^{n+1}\right] & \ldots & \frac{n-\alpha_1+3}{2^{n+1}(n+1)!}\frac{d^{\tau}}{dy^{\tau}}\left[(y^2-1)^{n+1}\right]\\
    {a}^{(2)}_{1,1}(y) & {a}^{(2)}_{1,2}(y)  & \ldots & {a}^{(2)}_{1,\alpha_1}(y)\\
    {a}^{(3)}_{2,1}(y) & {a}^{(3)}_{2,2}(y)  & \ldots & {a}^{(3)}_{2,\alpha_1}(y)\\
    \vdots & \vdots &  & \vdots\\
    {a}^{(\alpha_1-1)}_{\alpha_1-2,1}(y) & {a}^{(\alpha_1-1)}_{\alpha_1-2,2}(y)  & \ldots & {a}^{(\alpha_1-1)}_{\alpha_1-2,\alpha_1}(y)\\
\end{bmatrix}
\end{align}
where 
\begin{align} \label{eq:entries_after_row_operations}
  a_{i,j}^{(i+1)}=\begin{cases}\frac{n^{2}+o\left(n^{2}\right)}{2^{n+2}(n+2)!}\frac{d^{n+j-1}}{dy^{n+j-1}}\left[(y^2-1)^{n+2}\right]
 +\frac{1}{2^{n+1}(n+1)!}\frac{d^{n+j-1}}{dy^{n+j-1}}\left[(y^2-1)^{n+1}\right], & i=1\\
  \sum_{k=0}^{\floor{\frac{i+1}{2}}}\frac{c_k^{(i)}n^{i-2k+1} \left(1+o\left(1\right)\right)}{2^{n+i-k+1}(n+i-k+1)!}\frac{d^{n+j-1}}{dy^{n+j-1}}\left[(y^2-1)^{n+i-k+1}\right], &i=2,\ldots,\alpha_1-2
 \end{cases}
\end{align}
with fixed $c_k^{(i)}\in\mathbb{Z}^+$, and $c_0^{(i)}=1$, for all $k,i\in\mathbb{Z}$.  We then apply Lemma \ref{Lemma_legendre2} to the entries below the main diagonal, except for the terms generated by the sum in the second line of (\ref{eq:entries_after_row_operations}) when $j+k\geq i+2$, since they can be written in terms of derivatives of Legendre polynomials thanks to the Rodrigues' formula (\ref{eq:derivative_Legendre_binomial}). For the main diagonal and the upper-triangular entries, we also apply (\ref{eq:derivative_Legendre_binomial}), so that we obtain
\begin{equation}
  \begin{split}
 \mathbf{\hat{\Theta}}^{(\alpha_1-1)}_n(y)=
 \begin{bmatrix}
   d_1^n(y) & u_{1,1}^n(y) & \ldots & u_{1,\alpha_1-1}^n(y)  \\
    l_{1,1}^n(y) &  d_2^n(y) & \ddots & \vdots \\
      \vdots   & \ddots  & \ddots & u_{\alpha_1-1,\alpha_1-1}^n(y) \\
      l_{\alpha_1-1,1}^n(y) & \ldots  & l_{\alpha_1-1,\alpha_1-1}^n(y)  & d_{\alpha_1}^n(y)\\
\end{bmatrix}
\end{split}
\end{equation}  
where
\begin{equation}
  d_i^n(y)=\begin{cases}
  P_n(y), & i=1\\
 \sum_{k=0}^{\floor{\frac{i-1}{2}}}c_k^{(i-2)}\left(n^{i-2k-1}+o\left(n^{i-2k-1}\right)\right)\frac{d^{k}}{dy^{k}}P_{n+i-1-k}(y), & i=2,\ldots,\alpha_1,
  \end{cases}
\end{equation}
\begin{align}
l_{i,j}^n(y)=\begin{cases}
\sqrt{1-y^2}P_{n+2}^{1}(y) +P_{n+1}(y), & i=j=2 \\
\left(n^{-i+2j-2}+o\left(n^{-i+2j-2}\right)\right) \sum_{k=0}^{\floor{\frac{i}{2}}}c_k^{(i-1)}(1-y^2)^{(i-j-k+1)/2}P_{n+i-k}^{i-j-k+1}(y), &  j\leq \floor{\frac{i+3}{2}}\\
\left(n^{-i+2j-2}+o\left(n^{-i+2j-2}\right)\right)\sum_{k=0}^{i-j+1}c_k^{(i-1)}(1-y^2)^{(i-j-k+1)/2}P_{n+i-k}^{i-j-k+1}(y)\\
 +\sum_{k=i-j+2}^{\floor{\frac{i}{2}}}c_{k}^{(i-1)}\left(n^{i-2k}+o\left(n^{i-2k}\right)\right)\frac{d^{k+j-i-1}}{dy^{k+j-i-1}}P_{n+i-k}(y), & j\geq\floor{\frac{i+5}{2}},
\end{cases}
\end{align}
and
\begin{align}
u_{i,j}^n(y)=\begin{cases}
\frac{d^j}{dy^j}P_n(y) , & i=1\\
\sum_{k=0}^{\floor{\frac{i-1}{2}}}c_k^{(i-2)}\left(n^{i-2k-2}+o\left(n^{i-2k-2}\right)\right)\frac{d^{k+j-i+1}}{dy^{k+j-i+1}}P_{n+i-k-1}(y), & i\neq 1.
\end{cases}
\end{align}

As in the case $\alpha_1=3$ and $\alpha_2=0$, we further manipulate the entries to facilitate the application of Corollary \ref{corollary_legendre} and to make the entries of order $1$. We divide the $j$th column by $n^{j-1}(1-y^2)^{(\alpha_1-j)/2}$ for $j=1,\ldots,\alpha_1$, while multiplying the $i$th row by $(1-y^2)^{(\alpha_1-i)/2}$ for $i=1,\ldots,\alpha_1$. This gives $\det(\mathbf{E}_{\alpha_1}(y))=(y/n)^{-\alpha_1(\alpha_1-1)/2}\det({\mathbf{\tilde\Theta}}^{(\alpha_1-1)}_n(y))$ with
\begin{equation} \label{eq:eqv2}
  \mathbf{\tilde{\Theta}}^{(\alpha_1-1)}_n(y)=\begin{bmatrix}
   \tilde d_1^n(y) & \tilde u_{1,1}^n(y) & \ldots & \tilde u_{1,\alpha_1-1}^n(y)  \\
    \tilde l_{1,1}^n(y) & \tilde d_2^n(y) & \ddots & \vdots \\
      \vdots   & \ddots  & \ddots & \tilde u_{\alpha_1-1,\alpha_1-1}^n(y) \\
      \tilde l_{\alpha_1-1,1}^n(y) & \ldots  & \tilde l_{\alpha_1-1,\alpha_1-1}^n(y)  & \tilde d_{\alpha_1}^n(y)\\
\end{bmatrix}\end{equation}
where 
\begin{equation}
  \tilde d_i^n(y)=\begin{cases}
  P_n(y), & i=1\\
 \sum_{k=0}^{\floor{\frac{i-1}{2}}}c_k^{(i-2)}\left(n^{-2k}+o\left(n^{-2k}\right)\right)\frac{d^{k}}{dy^{k}}P_{n+i-1-k}(y), & i=2,\ldots,\alpha_1,
  \end{cases}
\end{equation}
\begin{align}
\tilde l_{i,j}^n(y)=\begin{cases}
n^{-1}P_{n+2}^{1}(y) +n^{-1}(1-y^2)^{-1/2}P_{n+1}(y), & i=j=2 \\
\left(n^{-i+j-1}+o\left(n^{-i+j-1}\right)\right) \sum_{k=0}^{\floor{\frac{i}{2}}}c_k^{(i-1)}(1-y^2)^{-k/2}P_{n+i-k}^{i-j-k+1}(y), &  j\leq \floor{\frac{i+3}{2}}\\
\left(n^{-i+j-1}+o\left(n^{-i+j-1}\right)\right)\sum_{k=0}^{i-j+1}c_k^{(i-1)}(1-y^2)^{-k/2}P_{n+i-k}^{i-j-k+1}(y)\\
 +\sum_{k=i-j+2}^{\floor{\frac{i}{2}}}c_{k}^{(i-1)}\frac{\left(n^{i-j-2k+1}+o\left(n^{i-j-2k+1}\right)\right)}{(1-y^2)^{(i-j+1)/2}}\frac{d^{k+j-i-1}}{dy^{k+j-i-1}}P_{n+i-k}(y), & j\geq\floor{\frac{i+5}{2}}
\end{cases}
\end{align}
and
\begin{align}
\tilde u_{i,j}^n(y)=\begin{cases}
\left(\frac{1}{n}\right)^j \left(1-y^2\right)^{j/2}\frac{d^j}{dy^j}P_n(y) , & i=1\\
\sum_{k=0}^{\floor{\frac{i-1}{2}}}c_k^{(i-2)}\left(n^{i-j-2k-1}+o\left(n^{i-j-2k-1}\right)\right)(1-y^2)^{(j-i+1)/2}\frac{d^{k+j-i+1}}{dy^{k+j-i+1}}P_{n+i-k-1}(y), & i\neq 1.
\end{cases}
\end{align}

Recall $x^\star=(1+x/n^2)/(1-x/n^2)$. We can now rewrite (\ref{eq:appendix_alpha2_0}) as
\begin{equation} \label{eq:equiv_det2}
  \begin{split}
  &F_{\phi_n}(x/n^2)=1-\frac{(1-x/n^2)^{\alpha_1 n+n^2+\alpha_1(\alpha_1-1)/2}}{(1+x/n^2)^{\alpha_1(\alpha_1-1)/2}}\frac{\det\left({\mathbf{\tilde\Theta}}^{(\alpha_1-1)}_n(x^\star)\right)}{\det\left({\mathbf{\tilde\Theta}}^{(\alpha_1-1)}_n(1)\right)}
  \end{split}
\end{equation}
and take the $n\rightarrow\infty$ limit. Specifically, applying Corollary \ref{corollary_legendre} to the entries of $\mathbf{\tilde\Theta}^{(\alpha_1-1)}_n(x^\star)$, and recalling that $1-y^2=-4x/n^2$, we obtain $\lim_{n\rightarrow\infty}\det({\mathbf{\tilde\Theta}}^{(\alpha_1-1)}_n(x^\star))=\det(\mathbf{L}^{(0)}(x))$ with
\begin{equation}
  \begin{split}
  \mathbf{L}^{(0)}(x)=\begin{bmatrix}
    \tilde d_1^\infty(x) & \tilde u_{1,1}^\infty(x) & \ldots & \tilde u_{1,\alpha_1-1}^\infty(x)  \\
    \tilde l_{1,1}^\infty(x) & \tilde d_2^\infty(x) & \ddots & \vdots \\
      \vdots   & \ddots  & \ddots & \tilde u_{\alpha_1-1,\alpha_1-1}^\infty(x) \\
      \tilde l_{\alpha_1-1,1}^\infty(x) & \ldots  & \tilde l_{\alpha_1-1,\alpha_1-1}^\infty(x)  & \tilde d_{\alpha_1}^\infty(x)\\
\end{bmatrix}
\end{split}
\end{equation}
where
\begin{align} \label{eq:di_inf}
 \tilde d_i^\infty(x)=\ &
 I_0(\sqrt{4x})+\sum_{k=1}^{\floor{\frac{i-1}{2}}}c_k^{(i-2)}(4x)^{-k/2}I_k(\sqrt{4x}),\\
 \label{eq:li_inf}
\tilde l_{i,j}^\infty(x)=\ &
(-\iota)^{i-j+1}I_{i-j+1}(\sqrt{4x})+ (-\iota)^{i-j+1}\sum_{k=1}^{\floor{\frac{i}{2}}}c_k^{(i-1)}(4x)^{-k/2}I_{i-j-k+1}(\sqrt{4x}),\\
 \label{eq:ui_inf}
\tilde u_{i,j}^\infty(x)=\ &
\iota^{j-i+1}I_{j-i+1}(\sqrt{4x})+\iota^{j-i+1}\sum_{k=1}^{\floor{\frac{i-1}{2}}}c_k^{(i-2)}(4x)^{-k/2}I_{k+j-i+1}(\sqrt{4x}),
\end{align}
since $I_m(z)=I_{-m}(z)$ for all $m\in\mathbb{N}$.

Now, we divide the $j$th column by $(-\iota)^{\alpha_1-j}$ for $j=1,\ldots,\alpha_1$ and multiply the $i$th row by $(-\iota)^{\alpha_1-i}$ for $i=1,\ldots,\alpha_1$, so that all complex constants vanish, just as in the previous subsection. We also perform row operations to get rid of the sums in (\ref{eq:di_inf})-(\ref{eq:ui_inf}), recalling that $I_m(z)=I_{-m}(z)$ for all $m\in\mathbb{N}$. In the first iteration, we manipulate the third and fourth rows. We subtract the second row scaled by $(4x)^{-1/2}$ from the third row. Then, we subtract the third row scaled by $(4x)^{-1/2}$ from the fourth row. In the second iteration, we manipulate the fifth and the sixth rows similarly. We repeat this procedure for a total of $\floor{(\alpha_1-1)/2}$ iterations, so that we can write the $n\rightarrow\infty$ limit of (\ref{eq:equiv_det2}) as
\begin{equation} \label{eq:appendix_after_asym2}
  \lim_{n\rightarrow\infty}F_{\phi_n}(x/n^2)=1-e^{-x}\frac{\det\left(\mathbf{L}^{(\floor{(\alpha_1-1)/2})}(x)\right)}{\det\left(\mathbf{L}^{(\floor{(\alpha_1-1)/2})}(0)\right)}
\end{equation}
where we have used the limit definition of the exponential and 
\begin{align}
  \det\left(\mathbf{L}^{(\floor{(\alpha_1-1)/2})}(x)\right)=\det\left[I_{i-j}(\sqrt{4x})\right]_{i=1,\ldots,\alpha_1}.
\end{align}
Considering that $I_k(0)=0$ for $k\in\mathbb{Z}^+$ and $I_0(0)=1$, explicit computation of the denominator in (\ref{eq:appendix_after_asym2}) gives $1$. Hence, we have proved the result for arbitrary $\alpha_1$ and $\alpha_2=0$. 

\subsection{Extension for arbitrary $\alpha_2$} \label{section:extension_arbitrary_alpha2}
Now consider the case $\alpha_2\neq 0$. Substituting (\ref{eq:g}) and (\ref{eq:legendre_minus_argument}) for the entries of the determinants $\mathbf{E}_{\alpha}(y)$ in (\ref{eq:H}) along with some algebraic simplifications, we obtain 
\begin{equation} \label{eq:smallest_xi}
   F_{n^2\phi_n}(x)=F_{\phi_n}(x/n^2)=1-(1-x/n^2)^{n^2+n\alpha_1+n\alpha_2+\alpha_1\alpha_2}\frac{\det\left(\mathbf{\Xi}^{(0)}_n(x^\star)\right)}{\det\left(\mathbf{\Xi}^{(0)}_n(1)\right)}
\end{equation}
with
\begin{equation} \label{eq:Xi_0}
  \begin{split}
  &\mathbf{\Xi}^{(0)}_n(y)= \begin{bmatrix}
    \mathbf{A}^{(0)}(y)  &  \mathbf{B}^{(0)}
\end{bmatrix}
\end{split}
\end{equation}
where $\mathbf{A}^{(0)}(y)$ is a $(\alpha_1+\alpha_2)\times\alpha_1$ matrix with entries
\begin{align}
  \left[\mathbf{A}^{(0)}(y)\right]_{ij}=(-1)^{i+j}\frac{d^{j-1}}{dy^{j-1}}P_{n+i-1}(y)
\end{align}
and $\mathbf{B}^{(0)}$ is a $(\alpha_1+\alpha_2)\times\alpha_2$ matrix with entries
\begin{align}
  \left[\mathbf{B}^{(0)}\right]_{ij}=(n+i-j+1)_{2j-2}.
\end{align}
 In the following, we apply a set of iterative operations to $\mathbf{\Xi}^{(0)}_n(y)$ to reduce the case $\alpha_2\neq 0$ to the case $\alpha_2=0$ when $n\rightarrow\infty$.   In the first iteration, we take advantage of the fact that the entries of the $(\alpha_1+1)$th column of $\mathbf{\Xi}^{(0)}_n(y)$ are all ones to reduce the dimension of the determinant by one. We successively subtract the $(k+1)$th row of $\mathbf{\Xi}^{(0)}_n(y)$ from the $k$th row, for $k=1,\ldots,\alpha_1+\alpha_2-1$, to make zero all the entries of the $(\alpha_1+1)$th column of $\mathbf{\Xi}^{(0)}_n(y)$ except that of the last row. We then simplify the modified entries beyond the $(\alpha_1+1)$th column with the help of Corollary \ref{corollary_property2} and the second line of (\ref{eq:g}), i.e.,
\begin{equation}
  2(m+1)(n-m+1)_{2m+1}=(n-m+1)_{2m+2}-(n-m)_{2m+2}.
\end{equation}
After this set of operations, the entries of that $(\alpha_1+1)$th column become all zero, except for that in the last row, which is not altered. We expand the determinant along this column to obtain $\det(\mathbf{\Xi}^{(0)}_n(y))=2^{-\alpha_2+2}\alpha_2^{-1}(\alpha_2-1)^{-1}\det(\mathbf{\Xi}^{(1)}_n(y))$ with
\begin{equation} \label{eq:Xi_1}
  \begin{split}
  &\mathbf{\Xi}^{(1)}_n(y)= \begin{bmatrix}
    \mathbf{A}^{(1)}(y)  &  \mathbf{B}^{(1)}
\end{bmatrix}
\end{split}
\end{equation}
where $\mathbf{A}^{(1)}(y)$ is a $(\alpha_1+\alpha_2-1)\times\alpha_1$ matrix with entries
\begin{equation}
  [\mathbf{A}^{(1)}(y)]_{ij}=(-1)^{i+j}\frac{d^{j-1}}{dy^{j-1}}(P_{n+i-1}(y)+P_{n+i}(y))
\end{equation}
and $\mathbf{B}^{(1)}$ is a $(\alpha_1+\alpha_2-1)\times(\alpha_2-1)$ matrix with entries
\begin{equation}
  [\mathbf{B}^{(1)}]_{ij}=(n+i-j+1)_{2j-1}.
  \end{equation}

In the second iteration, we successively subtract the $(k+1)$th row of $\mathbf{\Xi}^{(1)}_n(y)$ scaled by $(n+k)/(n+k+1)$ from the $k$th row, for $k=1,\ldots, \alpha_1+\alpha_2-2$.
Then, the entries of the $(\alpha_1+1)$th  column of $\mathbf{\Xi}^{(1)}_n(y)$ become all zeros except for that of the last row, which remains unchanged. We expand along this column to obtain $\det(\mathbf{\Xi}^{(0)}_n(y))=2^{-\alpha_2+2}((n+\alpha_1+\alpha_2-1)\alpha_2)^{-1}(\alpha_2-1)^{-1}\det(\mathbf{\Xi}^{(2)}_n(y))$ with
\begin{equation} \label{eq:Xi_2}
  \begin{split}
  &\mathbf{\Xi}^{(2)}_n(y)= \begin{bmatrix}
    \mathbf{A}^{(2)}(y)  &  \mathbf{B}^{(2)}
\end{bmatrix}
\end{split}
\end{equation}
where $\mathbf{A}^{(2)}(y)$ is a $(\alpha_1+\alpha_2-2)\times\alpha_1$ matrix with entries
\begin{equation}
  \left[\mathbf{A}^{(2)}(y)\right]_{ij}=
  (-1)^{i+j}\frac{d^{j-1}}{dy^{j-1}}\left(P_{n+i-1}(y)+P_{n+i}(y)+\frac{n+i}{n+i+1}(P_{n+i}(y)+P_{n+i+1}(y))\right)
\end{equation}
and $\mathbf{B}^{(2)}$ is a $(\alpha_1+\alpha_2-2)\times(\alpha_2-2)$ matrix with entries
\begin{equation}
  \left[\mathbf{B}^{(2)}\right]_{ij}=\frac{n+i}{n+i+1}(n+i-j+1)_{2j+1}-(n+i-j)_{2j+1}.
\end{equation}
In the following iterations, we repeat the same steps as in the second iteration, where we modify the scalings of the rows appropriately to make zero all the entries of the $(\alpha_1+1)$th column of $\mathbf{\Xi}^{(r)}_n(y)$ except for that of the last row. After a total of $\alpha_2$ iterations,  we rewrite (\ref{eq:smallest_xi}) as
\begin{equation} \label{eq:appendix_ratio}
  F_{\phi_n}(x/n^2)=1-(1-x/n^2)^{n^2+n\alpha_1+n\alpha_2+\alpha_1\alpha_2}\frac{\det\left(\mathbf{\Xi}^{(\alpha_2)}_n(x^\star)\right)}{\det\left(\mathbf{\Xi}^{(\alpha_2)}_n(1)\right)}
\end{equation}
where $\mathbf{\Xi}^{(\alpha_2)}_n(y)$ is a $\alpha_1\times\alpha_1$ matrix with entries
\begin{equation}
  \left[\mathbf{\Xi}^{(\alpha_2)}_n(y)\right]_{ij}=\sum_{k=0}^{\alpha_2}s_{k}^{(\alpha_2)}(n)\frac{d^{j-1}}{dy^{j-1}}P_{n+k+i-1}(y)
\end{equation}
with $s_k^{(\alpha_2)}(n)$ a ratio of polynomials in $n$ of the same order that does not depend on $j$. We do not need to explicitly define $s_k^{(\alpha_2)}(n)$ to complete the proof, since $s_k^{(\alpha_2)}(n)$ does not depend on $j$ and $\lim_{n\rightarrow\infty}\frac{d^{j-1}}{dy^{j-1}}P_{n+k+i-1}(y)$ does not depend on $k$ by virtue of Remark \ref{remark2}. Additionally, $s_k^{(\alpha_2)}(n)=\mathcal{O}(1)$ for all $k$, since $s_k^{(\alpha_2)}(n)$ is the result of multiplying and dividing entries of the $(\alpha_1+1)$th column of $\mathbf{\Xi}^{(r)}_n(y)$, which have the same order in $n$ for all $r$, and are exactly the same for $\mathbf{\Xi}^{(r)}_n(x^\star)$ and $\mathbf{\Xi}^{(r)}_n(1)$. Recalling Corollary \ref{corollary_legendre} and Remark \ref{remark2}, 
we then have
\begin{equation}
 \lim_{n\rightarrow\infty}n^{2-2j}\left[\mathbf{\Xi}^{(\alpha_2)}_n(x^\star)\right]_{ij}=\bar{s}\lim_{n\rightarrow\infty}n^{2-2j}\frac{d^{j-1}}{dy^{j-1}}P_{n+i-1}(y)\bigg\vert_{y=x^\star}
 \end{equation}
 where
 \begin{equation}
   \bar{s}=\sum_{k=0}^{\alpha_2}\lim_{n\rightarrow\infty}s_{k}^{(\alpha_2)}(n),
 \end{equation}
 which is $\mathcal{O}(1)$. Therefore,
 \begin{equation}
   \lim_{n\rightarrow\infty}n^{2-2j}\left[\mathbf{\Xi}^{(\alpha_2)}_n(x^\star)\right]_{ij}=\bar{s}\lim_{n\rightarrow\infty}n^{2-2j}\left[\mathbf{E}_{\alpha_1}(x^\star)\right]_{ij}.
    \end{equation}
We also have
\begin{equation}
   \lim_{n\rightarrow\infty}n^{2-2j}\left[\mathbf{\Xi}^{(\alpha_2)}_n(1)\right]_{ij}=\bar{s}\lim_{n\rightarrow\infty}n^{2-2j}\left[\mathbf{E}_{\alpha_1}(1)\right]_{ij}.
    \end{equation}
    This yields
   \begin{equation}
  \lim_{n\rightarrow\infty}\frac{\det\left(\mathbf{\Xi}^{(\alpha_2)}_n(x^\star)\right)}{\det\left(\mathbf{\Xi}^{(\alpha_2)}_n(1)\right)}=\lim_{n\rightarrow\infty}\frac{\det\left(\mathbf{E}_{\alpha_1}\left(x^\star\right)\right)}{\det\left(\mathbf{E}_{\alpha_1}(1)\right)},
\end{equation}
which concludes the proof.

\section{Proof of Proposition \ref{prop_correction1}} \label{sec:proofProp2}
We here prove the Proposition result for the different cases considered. We will only prove the result for the smallest eigenvalue distribution of $\mathbf{W}$ since, once this is established, the result for the largest eigenvalue follows immediately from Remark \ref{remark1}.
\subsection{Case $\alpha_1=0$ and arbitrary $\alpha_2$}

The case $\alpha_1=0$ is straightforward. From Corollary \ref{corollary_2}, we have
\begin{equation}
F_{n^2\phi_n}(x)=1-\left(1-\frac{x}{n^2}\right)^{n^2+n\alpha_2} ,
\end{equation}
and the result follows upon noting that, for fixed $\beta, \gamma$ (independent of $n$),
 \begin{equation} \label{eq:correction_binomial}
        \left(1-\frac{x}{n^2}\right)^{n^2+\beta n+\gamma}=e^{-x}-\frac{\beta x}{n}e^{-x}+\mathcal{O}\left(\frac{1}{n^2}\right).
\end{equation}

\subsection{Case $\alpha_1=1$ and arbitrary $\alpha_2$}
  The case $\alpha_1=1$ is more complicated, and for this we apply some results from Section \ref{section:extension_arbitrary_alpha2}; for consistency, we will use the same notation as in that section. Specifically, we will use (\ref{eq:appendix_ratio}) to write the distribution of the scaled smallest eigenvalue as
  \begin{equation} \label{eq:ratio_alpha1_1_arbitrary_alpha2}
      F_{n^2\phi_n}(x) = 1-\left(1-\frac{x}{n^2}\right)^{n^2+(1+\alpha_2)n+\alpha_2}\frac{\sum_{k=0}^{\alpha_2}s_k^{(\alpha_2)}(n)P_{n+k}\left(\frac{1+x/n^2}{1-x/n^2}\right)}{\sum_{k=0}^{\alpha_2}s_k^{(\alpha_2)}(n)},
  \end{equation}
along with the large-$n$ behaviour (including finite-$n$ corrections) of Legendre polynomials (Lemma \ref{Lemma_legendre_correction}) and that of the coefficients $s_k^{(\alpha_2)}(n)$. 

Since computing $s_k^{(\alpha_2)}$ for arbitrary $\alpha_2$ is not easy,
let us first explicitly compute them for the example case $\alpha_1=1$ and $\alpha_2=3$, in order to shed light on the properties required to prove the general result. To obtain (\ref{eq:ratio_alpha1_1_arbitrary_alpha2}), we start from (\ref{eq:smallest_xi}) where, for $\alpha_1=1$ and $\alpha_2=3$,
  \begin{equation} \label{eq:proof_proposition_alpha1_1_alpha2_arb}
  \begin{split}
      F_{n^2\phi_n}(x)=\ &1-\left(1-\frac{x}{n^2}\right)^{n^2+4n+3}\frac{\det\left(\mathbf{\Xi}_n^{(0)}\left(\frac{1+x/n^2}{1-x/n^2}\right)\right)}{\det\left(\mathbf{\Xi}_n^{(0)}(1)\right)}
    \end{split}
    \end{equation}
    with
    \begin{equation} \label{eq:Xi_0}
        \mathbf{\Xi}_n^{(0)}(y)=\begin{bmatrix}
    P_n\left(y\right) & 1 & (n)_2 & (n-1)_4\\
   -P_{n+1}\left(y\right)  & 1 & (n+1)_2 & (n)_4\\
    P_{n+2}\left(y\right) & 1 & (n+2)_2 & (n+1)_4\\
    -P_{n+3}\left(y\right) & 1 & (n+3)_2 & (n+2)_4\\
    \end{bmatrix}.
    \end{equation}
Following the steps described in Section \ref{section:extension_arbitrary_alpha2}, we apply a set of iterative operations to $\det(\mathbf{\Xi}^{(0)}_n(y))$, reducing successively the matrix dimension.  In the first iteration, we obtain $\det(\mathbf{\Xi}^{(0)}_n(y))=12^{-1}\det(\mathbf{\Xi}^{(1)}_n(y))$ with
  \begin{equation} \label{eq:Xi_1}
      \mathbf{\Xi}_n^{(1)}(y)=\begin{bmatrix}
    P_n\left(y\right)+P_{n+1}\left(y\right)  & n+1 & n(n+1)(n+2)\\
   -P_{n+1}\left(y\right)- P_{n+2}\left(y\right) & n+2 & (n+1)(n+2)(n+3)\\
    P_{n+2}\left(y\right)+ P_{n+3}\left(y\right) &  n+3 & (n+2)(n+3)(n+4)\\
    \end{bmatrix} ;
  \end{equation}
in the second iteration, we obtain $\det(\mathbf{\Xi}^{(0)}_n(y))=(12(n+3))^{-1}\det(\mathbf{\Xi}^{(2)}_n(y))$ with
  \begin{equation} \label{eq:Xi_2}
      \mathbf{\Xi}_n^{(2)}(y)=\begin{bmatrix}
    P_n\left(y\right)+P_{n+1}\left(y\right)+\frac{n+1}{n+2}\left(P_{n+1}(y)+P_{n+2}(y)\right)   & (n+1)(2n+3)\\
   -P_{n+1}\left(y\right)- P_{n+2}\left(y\right)-\frac{n+2}{n+3}\left(P_{n+2}(y)+P_{n+3}(y)\right)  & (n+2)(2n+5)\\
    \end{bmatrix} ;
  \end{equation}
  while after the last iteration, we obtain $\det(\mathbf{\Xi}^{(0)}_n(y))=(12(n+3)(n+2)(2n+5))^{-1}\det(\mathbf{\Xi}^{(3)}_n(y))$ with
  \begin{equation} \label{eq:Xi_3}
  \begin{split}
      \mathbf{\Xi}_n^{(3)}(y)=\ &
    P_n\left(y\right)+P_{n+1}\left(y\right)+\frac{n+1}{n+2}\left(P_{n+1}(y)+P_{n+2}(y)\right)\\
    &+\frac{(n+1)(2n+3)}{(n+2)(2n+5)}\left(P_{n+1}\left(y\right)+ P_{n+2}\left(y\right)+\frac{n+2}{n+3}\left(P_{n+2}(y)+P_{n+3}(y)\right)\right).
    \end{split}
  \end{equation}
  Observing each element of the first column of matrices $\mathbf{\Xi}_n^{(t)}(y)$, $t=0,1,2,3$, it is clear that the number of Legendre polynomial terms is doubled in each iteration, if one does not aggregate polynomials of the same degree. After aggregating polynomials of the same degree, we write
  \begin{equation} \label{eq:Xi_3_alt}
  \begin{split}
      \mathbf{\Xi}_n^{(3)}(y)=\sum_{k=0}^{3}s_k^{(3)}(n)P_{n+k}(y)
    \end{split}
  \end{equation}
  in agreement with (\ref{eq:ratio_alpha1_1_arbitrary_alpha2}), where
 \begin{align}
     s_0^{(3)}(n)=\ &1,\\
     s_1^{(3)}(n)=\ &1+\frac{n+1}{n+2}+\frac{(n+1)(2n+3)}{(n+2)(2n+5)},\\
     s_2^{(3)}(n)=\ &\frac{n+1}{n+2}+\frac{(n+1)(2n+3)}{(n+2)(2n+5)}+\frac{(n+1)(n+2)(2n+3)}{(n+2)(n+3)(2n+5)},\\
     s_3^{(3)}(n)=\ &\frac{(n+1)(n+2)(2n+3)}{(n+2)(n+3)(2n+5)}.
 \end{align}
 Note that these polynomial ratios can be expanded as $1+\mathcal{O}\left(1/n\right)$ since, for $i \in \mathbb{N}$,
  \begin{align}
     &\frac{n+i}{n+i+1}=1-\frac{1}{n}+\mathcal{O}\left(\frac{1}{n^2}\right),\\
     &\frac{2(n+i)+1}{2(n+i+1)+1}=1-\frac{1}{n}+\mathcal{O}\left(\frac{1}{n^2}\right).
 \end{align}
 From these expansions, we can see that
  \begin{align}
      s_1^{(3)}(n)=\ &3-\frac{3}{n}+\mathcal{O}\left(\frac{1}{n^2}\right),\\
      s_2^{(3)}(n)=\ &3-\frac{6}{n}+\mathcal{O}\left(\frac{1}{n^2}\right),\\
      s_3^{(3)}(n)=\ &1-\frac{3}{n}+\mathcal{O}\left(\frac{1}{n^2}\right),
 \end{align}
 and we generally write 
 \begin{equation}
     s_k^{(3)}(n)=a_k^{(3)}-\frac{b_k^{(3)}}{n}+\mathcal{O}\left(\frac{1}{n^2}\right),
 \end{equation}
 where $a_k^{(3)}$ is equal to the number of aggregated polynomial terms of degree $n+k$. It then follows that $\sum_{k=0}^{3}a_k^{(3)}$ is the total number of aggregated polynomial terms; indeed, we see that $\sum_{k=0}^{3}a_k^{(3)}=2^3$, consistent with the fact that the number of terms is doubled in each iteration. 
 
 Applying Lemma \ref{Lemma_legendre_correction} to (\ref{eq:Xi_3_alt}), we obtain
 \begin{equation} \label{eq:apply_Lemma3_Xi_3}
 \begin{split}
     \sum_{k=0}^{3}s_k^{(3)}(n)P_{n+k}(y) 
     =\ &\bar{a}^{(3)} I_0(\sqrt{4x})-\frac{\bar{b}^{(3)}}{n}I_0(\sqrt{4x})+\bar{c}^{(3)}\frac{\sqrt{x}}{n}I_1(\sqrt{4x})+\mathcal{O}\left(\frac{1}{n^2}\right)
     \end{split}
 \end{equation}
 where
 \begin{align}
     &\bar{a}^{(3)}=\sum_{k=0}^{3}a^{(3)}_k=8, \label{eq:sum_ak}\\
     &\bar{b}^{(3)}=\sum_{k=0}^{3}b_k^{(3)}=12,\\
     &\bar{c}^{(3)}=\sum_{k=0}^{3}a_k^{(3)}(1+2k)=32 \label{eq:sum_ak_1k} \; .
 \end{align}
Note also that evaluating (\ref{eq:apply_Lemma3_Xi_3}) at $y=\frac{1+x/n^2}{1-x/n^2}=1$ (i.e., at $x=0$) yields 
 \begin{equation} \label{eq:apply_Lemma3_Xi_3_y1}
 \sum_{k=0}^{3}s_k^{(3)}(n) 
     =\bar{a}^{(3)} -\frac{\bar{b}^{(3)}}{n} +\mathcal{O}\left(\frac{1}{n^2}\right).
 \end{equation}
 Therefore, we evaluate (\ref{eq:proof_proposition_alpha1_1_alpha2_arb}) (equivalently (\ref{eq:ratio_alpha1_1_arbitrary_alpha2})) as
  \begin{equation} \label{eq:end_correction_case_alpha1_1_arb_alpha2}
 \begin{split}
     F_{n^2\phi_n}(x)=
     \ &1-e^{-x}\left(1-\frac{4x}{n}\right)\frac{\bar{a}^{(3)} I_0(\sqrt{4x})-\frac{\bar{b}^{(3)}}{n}I_0(\sqrt{4x})+\bar{c}^{(3)}\frac{\sqrt{x}}{n}I_1(\sqrt{x})}{\bar{a}^{(3)} -\frac{\bar{b}^{(3)}}{n}}+\mathcal{O}\left(\frac{1}{n^2}\right)\\
      =\ &F_{\infty}^{(1)}(x)+\frac{e^{-x}}{n}\left(4x I_0(\sqrt{4x})-\frac{\bar{c}^{(3)}}{\bar{a}^{(3)}}\sqrt{x}I_1(\sqrt{4x})\right)+\mathcal{O}\left(\frac{1}{n^2}\right),\\
     =\ &F_{\infty}^{(1)}(x)+\frac{4x}{n}e^{-x}\left(I_0(\sqrt{4x})-\frac{1}{\sqrt{x}}I_1(\sqrt{4x})\right)+\mathcal{O}\left(\frac{1}{n^2}\right),
     \end{split}
 \end{equation}
 where the first equality follows from (\ref{eq:apply_Lemma3_Xi_3}), (\ref{eq:apply_Lemma3_Xi_3_y1}) and
  (\ref{eq:correction_binomial}).  The Proposition result (for $\alpha_1=1$ and $\alpha_2=3$) is obtained after noting that
 \begin{equation} \label{eq:recurrence_Bessel}
      I_{l+2}(z)=I_l(z)-\frac{2(l+1)}{z}I_{l+1}(z) \,, \quad  l\in\mathbb{Z} .
  \end{equation}
  From the second equality of (\ref{eq:end_correction_case_alpha1_1_arb_alpha2}), observe that, when considering the asymptotic distribution to order $\mathcal{O}\left(\frac{1}{n}\right)$, the quantity $\bar{b}^{(3)}$ has no effect (i.e., it drops out in the analysis). This will also occur in the general case of arbitrary $\alpha_2$, where we will only need to determine $\bar{a}^{(\alpha_2)}$ and $\bar{c}^{(\alpha_2)}$, which depend only on $a_k^{(\alpha_2)}$.
  
  Indeed, following the same steps of the previous example, we find that $\mathbf{\Xi}_n^{(\alpha_2)}(y) = \sum_{k=0}^{\alpha_2} s_k^{(\alpha_2)}(n) P_{n+k}(y)$ with $s_k^{(\alpha_2)}(n) = a_k^{(\alpha_2)} - b_k^{(\alpha_2)}/n + \mathcal{O}(1/n^2)$ and, using Lemma \ref{Lemma_legendre_correction} and (\ref{eq:correction_binomial}) in (\ref{eq:ratio_alpha1_1_arbitrary_alpha2}) we have, for arbitrary $\alpha_2$,
  \begin{equation} \label{eq:smallest_eig_proof_alpha1_1_alpha_2_arb}
\begin{split}
      F_{n^2\phi_n}(x) =\ &F_{\infty}^{(1)}(x)+\frac{e^{-x}}{n}\left((1+\alpha_2)x I_0(\sqrt{4x})-\frac{\bar{c}^{(\alpha_2)}}{\bar{a}^{(\alpha_2)}}\sqrt{x}I_1(\sqrt{4x})\right)+\mathcal{O}\left(\frac{1}{n^2}\right)
      \end{split}
  \end{equation}
   where 
   \begin{equation} \label{eq:bar_a_bar_c}
   \bar{a}^{(\alpha_2)}=\sum_{k=0}^{\alpha_2}a_k^{(\alpha_2)}, \quad \quad \bar{c}^{(\alpha_2)}=\sum_{k=0}^{\alpha_2}a_k^{(\alpha_2)}(1+2k).
  \end{equation}
   In light of (\ref{eq:recurrence_Bessel}), the proof will be complete if
  \begin{equation} \label{eq:induction_term}
      \frac{\bar{c}^{(\alpha_2)}}{\bar a^{(\alpha_2)}}=1+\alpha_2
  \end{equation}
holds. Let us now prove this equality.  

  From the previous example, we see that $a_k^{(\alpha_2)}$ equals the number of aggregated polynomial terms of degree $n+k$, after $\alpha_2$ iterations, and $\bar{a}^{(\alpha_2)}$ coincides with the total number of aggregated polynomial terms, irrespective of their degree (i.e., terms with the same degree are counted as different terms). We have also seen that the number of terms doubles after each iteration and, therefore,
  \begin{equation} \label{eq:number_terms}
      \bar{a}^{(\alpha_2)}=\sum_{k=0}^{\alpha_2}a_k^{(\alpha_2)}=2^{\alpha_2}.
  \end{equation}
  Furthermore, we have seen that, in each iteration, the additional polynomial terms have degrees increased by one (with respect to those terms in the previous iteration). This can be seen in the $(1,1)$ element of the matrices $\mathbf{\Xi}_n^{(t)}(y)$, $t=0,1,\ldots,\alpha_2$; see (\ref{eq:Xi_0})-(\ref{eq:Xi_3}) in the $\alpha_2=3$ example. From this, it becomes clear that 
  \begin{equation} \label{eq:recursion_ak}
      a_k^{(t+1)}=a_k^{(t)} + a_{k-1}^{(t)} 
  \end{equation}
  and, using this recursion, we can write
 \begin{align}
 \bar{c}^{(t+1)} = \sum_{k=0}^{t+1}a_k^{(t+1)}(1+2k) &= \sum_{k=0}^{t+1} (a_k^{(t)} + a_{k-1}^{(t)}) (1+2k) \\
 &= \sum_{k=0}^{t} a_k^{(t)} (1+2k) +  \sum_{k=0}^{t} a_k^{(t)} (1+2(k+1)) ,
  \end{align}
 where we have used the facts that $a_{-1}^{(t)}=0$ and $a_{t+1}^{(t)}=0$. From this, along with (\ref{eq:bar_a_bar_c}) and (\ref{eq:number_terms}), it is then clear that
 \begin{equation} \label{eq:induction_recursive}
       \bar{c}^{(t+1)}=2\bar{c}^{(t)}+2^{t+1}.
 \end{equation}

  Thanks to (\ref{eq:induction_recursive}), we now prove (\ref{eq:induction_term}) by induction. For $\alpha_1=1$ and $\alpha_2=1$, 
  
  \begin{equation} \label{induction1}
  \begin{split} 
      F_{n^2\phi_n}(x)=\ &1-\left(1-\frac{x}{n^2}\right)^{n^2+2n+1}\frac{\det\begin{bmatrix}
    P_n\left(\frac{1+x/n^2}{1-x/n^2}\right) & 1\\
   -P_{n+1}\left(\frac{1+x/n^2}{1-x/n^2}\right)  & 1
    \end{bmatrix}}{\det\begin{bmatrix}
    1 & 1\\
   -1  & 1
    \end{bmatrix}}\\
    =\ &1-\left(1-\frac{x}{n^2}\right)^{n^2+2n+1}\frac{P_n\left(\frac{1+x/n^2}{1-x/n^2}\right)+P_{n+1}\left(\frac{1+x/n^2}{1-x/n^2}\right)}{2}\\
    \end{split}
  \end{equation}
  where we identify that $s_0^{(1)}(n)=s_1^{(1)}(n)=1$ in light of (\ref{eq:ratio_alpha1_1_arbitrary_alpha2}). Then, $a_0^{(1)}=a_1^{(1)}=1$, which gives $\bar c^{(1)}=4$ and $\bar a^{(1)}$=2. Thus, $\bar c^{(\alpha_2)} / \bar a^{(\alpha_2)}=1+\alpha_2$ holds for $\alpha_2=1$. 
  

  For $\alpha_1=1$ and $\alpha_2=t$, suppose that
  \begin{equation}
      \frac{\bar c^{(t)}}{2^{t}}=1+t.
  \end{equation}
  
  Then, for $\alpha_2=t+1$,
  \begin{equation}
      \frac{\bar c^{(t+1)}}{2^{t+1}}=\frac{2\bar c^{(t)}+2^{t+1}}{2^{t+1}}=2+t
  \end{equation}
  where (\ref{eq:induction_recursive}) has been used. This proves (\ref{eq:induction_term}) and the result of Proposition \ref{prop_correction1} is obtained after applying (\ref{eq:induction_term}) and (\ref{eq:recurrence_Bessel}) to (\ref{eq:smallest_eig_proof_alpha1_1_alpha_2_arb}).
  
  \subsection{Case $\alpha_1=2$ and $\alpha_2=0$}
 For this case, the scaled smallest eigenvalue distribution is given by
  \begin{equation} \label{eq:proof_correction_alpha1_2}
  F_{n^2\phi_n}(x)=1-\left(1-\frac{x}{n^2}\right)^{n^2+2n}\frac{\det\begin{bmatrix}
    P_n\left(\frac{1+x/n^2}{1-x/n^2}\right) & \frac{dP_n(y)}{dy}\vert_{y=\frac{1+x/n^2}{1-x/n^2}} \\
  P_{n+1}\left(\frac{1+x/n^2}{1-x/n^2}\right)  & \frac{dP_{n+1}(y)}{dy}\vert_{y=\frac{1+x/n^2}{1-x/n^2}} \\
    \end{bmatrix}}{\det\begin{bmatrix}
     P_n\left(1\right) & \frac{dP_n(y)}{dy}\vert_{y=1} \\
  P_{n+1}\left(1\right)  & \frac{dP_{n+1}(y)}{dy}\vert_{y=1} \\
    \end{bmatrix}}.
    \end{equation}
    Performing the same row operations as in Section \ref{proof_3x3}, we rewrite (\ref{eq:proof_correction_alpha1_2}) in the form of (\ref{eq:equiv_det2_alpha3}), i.e.
    \begin{equation} 
  F_{n^2\phi_n}(x)=1-\frac{\left(1-\frac{x}{n^2}\right)^{n^2+2n+2}}{\left(1+x/n^2\right)^2}\frac{\det\begin{bmatrix}
    P_n\left(\frac{1+x/n^2}{1-x/n^2}\right) & -\frac{ P_n^1\left(\frac{1+x/n^2}{1-x/n^2}\right)}{n} \\
  \frac{P_{n+1}^1\left(\frac{1+x/n^2}{1-x/n^2}\right)}{n+1}  & \frac{n+1}{n}P_{n+1}\left(\frac{1+x/n^2}{1-x/n^2}\right) 
    \end{bmatrix}}{\det\begin{bmatrix}
     P_n\left(1\right) & -\frac{ P_{n}^1\left(1\right)}{n} \\
  \frac{P_{n+1}^1\left(1\right)}{n+1}  & \frac{n+1}{n}P_{n+1}\left(1\right)  \\
    \end{bmatrix}}.
    \end{equation}
  By virtue of Lemma \ref{Lemma_legendre_correction} and (\ref{eq:correction_binomial}), we obtain
  \begin{equation} \label{eq:case_alpha1_2_alpha1_0_after_manip}
  \begin{split}
      F_{n^2\phi_n}(x)=\ &1-e^{-x}\left(1-\frac{2x}{n}\right)\\
      &\times\frac{\det\begin{bmatrix}
    I_0(\sqrt{4x})+\frac{\sqrt{x}}{n}I_1(\sqrt{4x}) & \iota I_1(\sqrt{4x})+\frac{\iota\sqrt{x}}{n} I_0(\sqrt{4x})\\
  -\left(1-\frac{1}{n}\right)\left(\iota I_1(\sqrt{4x})-\frac{\iota 3\sqrt{x}}{n}I_0(\sqrt{4x})\right)  & \left(1+\frac{1}{n}\right)\left(I_0(\sqrt{4x})+\frac{3\sqrt{x}}{n}I_1(\sqrt{4x})\right) 
    \end{bmatrix}}{\det\begin{bmatrix}
     1 & 0 \\
  0  & 1+\frac{1}{n}  \\
    \end{bmatrix}}+\mathcal{O}\left(\frac{1}{n^2}\right)
    \end{split}
  \end{equation}
  since
  \begin{equation}
      \frac{1}{\left(1+\frac{x}{n^2}\right)^2}=1-\frac{2x}{n^2}+\mathcal{O}\left(\frac{1}{n^4}\right).
  \end{equation}
  Noticing that
  \begin{equation}
      \frac{1}{\det\begin{bmatrix}
     1 & 0 \\
  0  & 1+\frac{1}{n}  \\
    \end{bmatrix}}=1-\frac{1}{n}+\mathcal{O}\left(\frac{1}{n^2}\right),
  \end{equation}
  we develop the numerator determinant in (\ref{eq:case_alpha1_2_alpha1_0_after_manip}) and we obtain the result after some manipulations.
  
  \subsection{Case $\alpha_1=2$ and $\alpha_2=1$}
 For this case, the scaled smallest eigenvalue distribution is given by
  \begin{equation} \label{eq:proof_correction_alpha1_2_alpha_2_1}
  F_{n^2\phi_n}(x)=1-\left(1-\frac{x}{n^2}\right)^{n^2+3n+2}\frac{\det\begin{bmatrix}
    P_n\left(\frac{1+x/n^2}{1-x/n^2}\right) & -\frac{dP_n(y)}{dy}\vert_{y=\frac{1+x/n^2}{1-x/n^2}} & 1\\
  -P_{n+1}\left(\frac{1+x/n^2}{1-x/n^2}\right)  & \frac{dP_{n+1}(y)}{dy}\vert_{y=\frac{1+x/n^2}{1-x/n^2}} & 1 \\
  P_{n+2}\left(\frac{1+x/n^2}{1-x/n^2}\right)  & -\frac{dP_{n+2}(y)}{dy}\vert_{y=\frac{1+x/n^2}{1-x/n^2}} & 1 \\
    \end{bmatrix}}{\det\begin{bmatrix}
     P_n\left(1\right) & -\frac{dP_n(y)}{dy}\vert_{y=1} & 1 \\
  -P_{n+1}\left(1\right)  & \frac{dP_{n+1}(y)}{dy}\vert_{y=1} & 1 \\
  P_{n+2}\left(1\right)  & -\frac{dP_{n+2}(y)}{dy}\vert_{y=1} & 1 \\
    \end{bmatrix}}.
    \end{equation}
    Performing the same row operations as in Section \ref{section:extension_arbitrary_alpha2}, we rewrite (\ref{eq:proof_correction_alpha1_2_alpha_2_1}) in the form of (\ref{eq:appendix_ratio}), i.e.,
    \begin{equation} 
  F_{n^2\phi_n}(x)=1-(1-x/n^2)^{n^2+3n+2}\frac{\det\left(\mathbf{\Xi}^{(1)}_n\left(\frac{1+x/n^2}{1-x/n^2}\right)\right)}{\det\left(\mathbf{\Xi}^{(1)}_n(1)\right)}
\end{equation}
where
    \begin{equation} 
 \mathbf{\Xi}^{(1)}_n(y)=\begin{bmatrix}
    P_n\left(y\right)+P_{n+1}\left(y\right) & \frac{dP_n(y)}{dy}+\frac{dP_{n+1}(y)}{dy} \\
  P_{n+1}\left(y\right)+P_{n+2}\left(y\right)  & \frac{dP_{n+1}(y)}{dy}+\frac{dP_{n+2}(y)}{dy}  \\
    \end{bmatrix}.
    \end{equation}
    Now, we perform the manipulations described in Section \ref{proof_3x3}, and we write
    \begin{equation} \label{eq:proof_corr_ratio_theta1}
  F_{n^2\phi_n}(x)=1-\frac{(1-x/n^2)^{n^2+3n+3}}{(1+x/n^2)}\frac{\det\left(\mathbf{\tilde\Theta}^{(1)}_n\left(\frac{1+x/n^2}{1-x/n^2}\right)\right)}{\det\left(\mathbf{\tilde\Theta}^{(1)}_n(1)\right)}
\end{equation}
where
    \begin{equation} \label{eq:proof_corr_theta1}
 \mathbf{\tilde\Theta}^{(1)}_n(y)=\begin{bmatrix}
    P_n\left(y\right)+P_{n+1}\left(y\right) & \frac{\sqrt{1-y^2}}{n}\left(\frac{dP_n(y)}{dy}+\frac{dP_{n+1}(y)}{dy}\right) \\
  \frac{P_{n+1}^1\left(y\right)}{n+1}+\frac{P_{n+2}^1\left(y\right)}{n+2}  & \frac{n+1}{n}P_{n+1}(y)+\frac{n+2}{n}P_{n+2}(y)  \\
    \end{bmatrix}.
    \end{equation}
    When applying Lemma \ref{Lemma_legendre_correction} to (\ref{eq:proof_corr_theta1}), we obtain
    \begin{equation}  \label{eq:proof_corr_theta1_after_lemma}
 \mathbf{\tilde\Theta}^{(1)}_n(y)=\begin{bmatrix}
    2I_0(\sqrt{4x})+\frac{4\sqrt{x}}{n}I_{1}(\sqrt{4x})+\mathcal{O}\left(\frac{1}{n^2}\right) & 2I_1(\sqrt{4x})+\frac{4\sqrt{x}}{n}I_{0}(\sqrt{4x})+\mathcal{O}\left(\frac{1}{n^2}\right) \\
  \left(2-\frac{3}{n}\right)I_1(\sqrt{4x})+\frac{8\sqrt{x}}{n}I_{0}(\sqrt{4x})+\mathcal{O}\left(\frac{1}{n^2}\right)  & \left(2+\frac{3}{n}\right)I_0(\sqrt{4x})+\frac{8\sqrt{x}}{n}I_{1}(\sqrt{4x})+\mathcal{O}\left(\frac{1}{n^2}\right)  \\
    \end{bmatrix}.
    \end{equation}
    We then apply some row operations to (\ref{eq:proof_corr_theta1_after_lemma}) so that the determinant remains unaltered. Specifically, we scale the second row by $2\sqrt{x}/n$ and we subtract it from the first row. Then, we scale the first row by $4\sqrt{x}/{n}$ and we subtract it from the second row. Therefore, we evaluate (\ref{eq:proof_corr_ratio_theta1}) as
    \begin{equation} \label{eq:proof_corr_ratio_thea1_after_lemma}
        F_{n^2\phi_n}(x)=1-e^{-x}\left(1-\frac{3x}{n}\right)\frac{\det\begin{bmatrix}
    2I_0(\sqrt{4x}) & 2I_1(\sqrt{4x})\\
  \left(2-\frac{3}{n}\right)I_1(\sqrt{4x})  & \left(2+\frac{3}{n}\right)I_0(\sqrt{4x})  \\
    \end{bmatrix}}{\det\begin{bmatrix}
    2 & 0\\
  0  & \left(2+\frac{3}{n}\right)  \\
    \end{bmatrix}}+\mathcal{O}\left(\frac{1}{n^2}\right)
    \end{equation}
    where (\ref{eq:correction_binomial}) has been used. Since 
    \begin{equation}
        \frac{1}{\det\begin{bmatrix}
    2 & 0\\
  0  & \left(2+\frac{3}{n}\right)  \\
    \end{bmatrix}}=\left(\frac{1}{4}-\frac{3}{8n}\right)+\mathcal{O}\left(\frac{1}{n^2}\right),
    \end{equation}
    we have the result when developing the numerator determinant in (\ref{eq:proof_corr_ratio_thea1_after_lemma}), performing some simplifications and applying (\ref{eq:recurrence_Bessel}).
    
    \subsection{Case $\alpha_1=\alpha_2=2$}
For this case, the scaled smallest eigenvalue distribution is given by
  \begin{equation} \label{eq:proof_correction_alpha1_2_alpha_2_2}
  F_{n^2\phi_n}(x)=1-\left(1-\frac{x}{n^2}\right)^{n^2+4n+4}\frac{\det\begin{bmatrix}
    P_n\left(\frac{1+x/n^2}{1-x/n^2}\right) & -\frac{dP_n(y)}{dy}\vert_{y=\frac{1+x/n^2}{1-x/n^2}} & 1 & (n)_2\\
  -P_{n+1}\left(\frac{1+x/n^2}{1-x/n^2}\right)  & \frac{dP_{n+1}(y)}{dy}\vert_{y=\frac{1+x/n^2}{1-x/n^2}} & 1 & (n+1)_2\\
  P_{n+2}\left(\frac{1+x/n^2}{1-x/n^2}\right)  & -\frac{dP_{n+2}(y)}{dy}\vert_{y=\frac{1+x/n^2}{1-x/n^2}} & 1 & (n+2)_2\\
 -P_{n+3}\left(\frac{1+x/n^2}{1-x/n^2}\right)  & \frac{dP_{n+3}(y)}{dy}\vert_{y=\frac{1+x/n^2}{1-x/n^2}} & 1 & (n+3)_2\\
    \end{bmatrix}}{\det\begin{bmatrix}
     P_n\left(1\right) & -\frac{dP_n(y)}{dy}\vert_{y=1} & 1 & (n)_2\\
  -P_{n+1}\left(1\right)  & \frac{dP_{n+1}(y)}{dy}\vert_{y=1} & 1 & (n+1)_2\\
  P_{n+2}\left(1\right)  & -\frac{dP_{n+2}(y)}{dy}\vert_{y=1} & 1 & (n+2)_2\\
  -P_{n+3}\left(1\right)  & \frac{dP_{n+3}(y)}{dy}\vert_{y=1} & 1 & (n+3)_2\\
    \end{bmatrix}}.
    \end{equation}
    Performing the same row operations as in Section \ref{section:extension_arbitrary_alpha2}, we rewrite (\ref{eq:proof_correction_alpha1_2_alpha_2_2}) in the form of (\ref{eq:appendix_ratio}), i.e.,
    \begin{equation} \label{eq:proof_corr_ratio_theta2}
  F_{n^2\phi_n}(x)=1-(1-x/n^2)^{n^2+4n+4}\frac{\det\left(\mathbf{\Xi}^{(2)}_n\left(\frac{1+x/n^2}{1-x/n^2}\right)\right)}{\det\left(\mathbf{\Xi}^{(2)}_n(1)\right)}
\end{equation}
where $\mathbf{\Xi}^{(2)}_n(y)$ is a $2\times 2$ matrix with entries
    \begin{equation} 
 \left[\mathbf{\Xi}^{(2)}_n(y)\right]=\frac{d^{j-1}}{d^{j-1}}\left[P_{n+i-1}(y)+P_{n+i}+\frac{n+i}{n+i+1}\left(P_{n+i}(y)+P_{n+i+1}(y)\right)\right].
    \end{equation}
As in Section \ref{proof_3x3}, we perform some row operations to $\mathbf{\Xi}^{(2)}_n(y)$ to facilitate the application of Lemma \ref{Lemma_legendre_correction}. Specifically, we scale the second row of $\mathbf{\Xi}^{(2)}_n(y)$ by $y$ and we subtract the first row. We also divide the $j$th column by $n^{j-1}(1-y^2)^{(2-j)/2}$ for $j=1,2$, and multiply the $i$th row by $(1-y^2)^{(2-i)/2}$ for $i=1,2$. We then rewrite (\ref{eq:proof_corr_ratio_theta2}) as 
\begin{equation} \label{eq:proof_corr_ratio_theta2_tilde}
  F_{n^2\phi_n}(x)=1-\frac{(1-x/n^2)^{n^2+4n+5}}{1+x/n^2}\frac{\det\left(\mathbf{\tilde\Xi}^{(2)}_n\left(\frac{1+x/n^2}{1-x/n^2}\right)\right)}{\det\left(\mathbf{\tilde\Xi}^{(2)}_n(1)\right)}
\end{equation}
where 
\begin{equation} 
 \mathbf{\tilde\Xi}^{(2)}_n(y)=\begin{bmatrix}
    e_{11}(y) & e_{12}(y) \\
  e_{21}(y)  & e_{22}(y)   \\
    \end{bmatrix}
    \end{equation}
    with
    \begin{align}
        e_{11}(y)=\ &P_n\left(y\right)+P_{n+1}\left(y\right)+\frac{n+1}{n+2}\left(P_{n+1}\left(y\right)+P_{n+2}\left(y\right)\right),\\
        e_{12}(y)=\ &\frac{\sqrt{1-y^2}}{n}\left(\frac{dP_n(y)}{dy}+\frac{dP_{n+1}(y)}{dy}+\frac{n+1}{n+2}\left(\frac{dP_{n+1}\left(y\right)}{dy}+\frac{dP_{n+2}\left(y\right)}{dy}\right)\right),\\
        e_{21}(y)=\ &\frac{P_{n+1}^{1}(y)}{n+1}+\frac{P_{n+2}^{1}(y)}{n+2}+\frac{P_{n+2}^{1}(y)}{n+3}+\frac{(1-y^2)^{-1/2}}{(n+2)(n+3)}(P_{n+1}(y)+P_{n+2}(y))+\frac{n+2}{(n+3)^2}P_{n+3}^1(y),\\
        e_{22}(y)=\ &\frac{n+1}{n}P_{n+1}(y)+\frac{n+2}{n}P_{n+2}(y)+\frac{(n+2)^2}{(n+3)n}P_{n+2}(y)+\frac{n+2}{n}P_{n+3}(y)\\
        &+\frac{1}{(n+2)(n+3)n}\left(\frac{dP_{n+1}(y)}{dy}+\frac{dP_{n+2}(y)}{dy}\right).
    \end{align}
    
    We then apply Lemma \ref{Lemma_legendre_correction} to the entries of $\mathbf{\tilde\Xi}^{(2)}_n(y)$, and expand the ratio of polynomials in $n$ to obtain, after aggregating terms,
    \begin{align}
        e_{11}(y)=\ &4I_0(\sqrt{4x})-2\frac{I_0(\sqrt{4x})}{n}+12\frac{\sqrt{x}}{n}I_1(\sqrt{4x})+\mathcal{O}\left(\frac{1}{n^2}\right),\\
        e_{12}(y)=\ &4\iota I_1(\sqrt{4x})-2\iota\frac{I_1(\sqrt{4x})}{n}+12\iota\frac{\sqrt{x}}{n}I_0(\sqrt{4x})+\mathcal{O}\left(\frac{1}{n^2}\right),\\
        e_{21}(y)=\ &-4\iota I_1(\sqrt{4x})+10\iota\frac{I_1(\sqrt{4x})}{n}-20\iota\frac{\sqrt{x}}{n}I_0(\sqrt{4x})-8\iota\frac{I_0(\sqrt{4x})}{\sqrt{4x}}+\mathcal{O}\left(\frac{1}{n^2}\right),\\
        e_{22}(y)=\ &4I_0(\sqrt{4x})+6\frac{I_0(\sqrt{4x})}{n}+20\frac{\sqrt{x}}{n}I_1(\sqrt{4x})+8\frac{I_1(\sqrt{4x})}{\sqrt{4x}}+\mathcal{O}\left(\frac{1}{n^2}\right).
    \end{align}
    
    Using these asymptotic expansions for the determinant entries in (\ref{eq:proof_corr_ratio_theta2_tilde}), we then compute those determinants, make some simplifications and obtain the result with the help of (\ref{eq:correction_binomial}) and (\ref{eq:recurrence_Bessel}), similarly as in the previous cases.

\bibliography{bibfile}

\end{document}